\documentclass[12pt]{article}
\usepackage{amssymb,slashed,latexsym,ifpdf,amsmath,amsthm}
\usepackage{nicefrac}
\newcommand{\ft}[2]{{\textstyle\frac{#1}{#2}}}

\def\Im{\mathop{\rm Im}\nolimits}

\def\rme{{\rm e}}
\def\rmi{{\rm i}}

\newcommand{\poinc}{\boxdot}   
\newsavebox{\uuunit}
\sbox{\uuunit}
    {\setlength{\unitlength}{0.825em}
     \begin{picture}(0.6,0.7)
        \thinlines
        \put(0,0){\line(1,0){0.5}}
        \put(0.15,0){\line(0,1){0.7}}
        \put(0.35,0){\line(0,1){0.8}}
       \multiput(0.3,0.8)(-0.04,-0.02){12}{\rule{0.5pt}{0.5pt}}
     \end {picture}}

\newcommand{\hc}{{\rm h.c.}}
\newcommand{\bbox}{\lower.2ex\hbox{$\Box$}}
\newcommand{\wcD}{\widehat{\cal D }}

\csname @addtoreset\endcsname{equation}{section}

\newcommand{\U}{\mathop{\rm {}U}}

\newif\ifpdf
\ifx\pdfoutput\undefined
   \pdffalse
   \usepackage{cite}
 \else
   \pdfoutput=1
   \pdftrue
  \usepackage[pdftex]{hyperref}
\fi
\usepackage{color}
  	\definecolor{antiquefuchsia}{rgb}{0.57, 0.36, 0.51}
 
\def\aD3{{\overline {\rm D3}}}
\def\be{\begin{equation}}
\def\ee{\end{equation}}
\def\ba{\begin{array}}
\def\ea{\end{array}}
\def\bea{\begin{eqnarray}}
\def\eea{\end{eqnarray}}

\def\a{{\alpha}}

\def\b{{\beta}}

\newcommand{\ca}{{\cal A}}
\newcommand{\cd}{{\cal D}}

\newcommand{\ck}{{\cal K}}

\def\be{\begin{equation}}
\def\ee{\end{equation}}
\def\bea{\begin{eqnarray}}
\def\eea{\end{eqnarray}}
\def\ba{\begin{array}}
\def\ea{\end{array}}
\def\bd{\begin{displaymath}}
\def\ed{\end{displaymath}}

\def\a{\alpha}
\def\b{\beta}

\def\d{\delta}
\def\e{\epsilon}           
  
\def\g{\gamma}

\def\l{\lambda}
\def\m{\mu}
\def\n{\nu}
\def\o{\omega}  

\def\G{\Gamma}

\def\O{\Omega}

\def\pa{\partial}                              
\def\>{\rangle} 
\def\<{\langle} 
\def\Dsl{D \hskip-.6em \raise1pt\hbox{$ / $ } }
\def\to{\rightarrow}
\def\pa{\partial}

\newcommand{\eps}{\epsilon}

\def\Kahler{K\"{a}hler~}
\def\Poincare{Poincar\'e~}
\def\susy{supersymmetry~}

\begin{document}

\begin{titlepage}
\vspace{.5cm}
\begin{center}
\baselineskip=16pt

\hskip 1cm

\vskip 0.8cm

{\huge {\bf Off-Shell Poincar\'{e} Supergravity}}

\

\

\

{\large  \bf Daniel Z.  Freedman$^{1,2}$},
{ \large  \bf Diederik Roest$^3$}, and   {\large  \bf Antoine Van Proeyen$^4$} \vskip 0.8cm
{\small\sl\noindent
$^1$ SITP and Department of Physics, Stanford University, Stanford, California
94305 USA \\\smallskip
$^2$ Center for Theoretical Physics and Department of Mathematics,\\ Massachusetts Institute of Technology, Cambridge, Massachusetts 02139, USA \\\smallskip
$^3$ Van Swinderen Institute for Particle Physics and Gravity,\\
University of Groningen, Nijenborgh 4, 9747 AG Groningen, The Netherlands\\\smallskip
$^4$ KU Leuven, Institute for Theoretical Physics,\\ Celestijnenlaan 200D, B-3001 Leuven,
Belgium}

\end{center}


\vskip 2cm
\begin{center}
{\bf Abstract}

\end{center}

 {\small We present the action and transformation rules of Poincar\'{e} supergravity coupled to chiral multiplets $(z^\alpha, \chi^\alpha, h^\alpha)$ with off-shell auxiliary fields. Starting from the geometric formulation of the superconformal theory with auxiliary fields, we derive the Poincar\'{e} counterpart by gauge-fixing the Weyl and chiral symmetry and $S$-supersymmetry. We show how this transition is facilitated by retaining explicit target-space covariance. Our results form a convenient starting point to study models with constrained superfields, including general matter-coupled de Sitter supergravity. }
   \vspace{2mm} \vfill \hrule width 3.cm
{\footnotesize \noindent E-mails:  dzf@math.mit.edu, d.roest@rug.nl, Antoine.VanProeyen@fys.kuleuven.be
 }
 \end{titlepage}
\addtocounter{page}{1}
 \tableofcontents{}
\newpage
\section{Introduction}

In this paper we reformulate ${\cal N}=1,~D=4$ supergravity theory coupled to chiral multiplets using ingredients that are manifestly covariant under complex diffeomorphisms of the K\"{a}hler target space. This  simplifies the action and transformation rules at the superconformal level  and streamlines the passage to the physical theory, which is invariant under local Poincar\'{e} supersymmetry. We explicitly retain the auxiliary fields of the chiral multiplets, keeping these off-shell. This formulation permits the construction of supergravity theories in which supersymmetry is realized nonlinearly due to constraints on superfields.

Nonlinear supersymmetry breaks the usual degeneracy between bosonic and fermionic degrees of freedom. It can be seen to follow from imposing supersymmetric constraints on the superfield, leading to constrained superfields. The present constructions include  a nilpotent field coupled to the supergravity multiplet \cite{Bergshoeff:2015tra, Hasegawa:2015bza} as well as physical chiral and vector multiplets \cite{Kallosh:2015sea, Kallosh:2015tea, Schillo:2015ssx}. Moreover, extensions have been proposed involving additional constrained superfields subject to orthogonal constraints \cite{Dall'Agata:2015zla, Ferrara:2015tyn}. An overview of such possibilities can be found in \cite{DallAgata:2015zxp, Ferrara:2016een}. Finally, one can construct de Sitter supergravity by imposing constraints involving the supergravity multiplet as well \cite{Delacretaz:2016nhw, Cribiori:2016qif}.

Given these prolific developments in constrained superfields, it is clearly advantageous to solve for any auxiliary fields at the latest possible stage, in order to allow for as many constraints. This is exactly what our current formulation provides, with the final \Poincare result containing the auxiliary fields $h^0$ and $h^\alpha$ of the supergravity and chiral multiplets. The latter correspond to the order parameters of supersymmetry breaking. In the case of linearly realized supersymmetry, these are taken as auxiliary fields of a non-constrained superfield and take their Gaussian values in terms of the physical components. In the case of a non-linear realization, however, the auxiliary field is a free parameter and instead some of the otherwise physical components are solved in terms of this parameter.

The proposed framework allows for arbitrary numbers of constrained and independent superfields. For instance, it can be used to analyze the situation with a number of nilpotent superfields in addition to independent chiral multiplets. Similarly, it allows for arbitrary couplings between the different types of fields.

An important role in our derivation is played by the covariance of supersymmetry transformations and auxiliary fields. We will employ a formulation in which these transform covariantly under reparametrizations of the target space spanned by the scalars $\phi$ of the theory. This formulation was emphasized very recently in \cite{Freedman:2016qnq}, where further details and motivation can be found. The key point is that symmetry transformations $\delta \chi$ (which can include Killing isometries, supersymmetry, ....) generically do not transform covariantly under reparametrizations. To remedy this situation, one can introduce covariant transformations and derivatives defined by
 \begin{align}
  \hat \delta \chi^i = \delta \chi^i + \Gamma^i_{jk} \chi^j \delta \phi^k \,, \quad
 \nabla_\mu \chi^i = \partial_\mu \chi^i + \Gamma^i_{jk} \chi^j \partial_\mu \phi^k \,. \label{covariantdef}
 \end{align}
These are the unique quantities that transform covariantly both under target space repara\-metrizations as well as other symmetries.

Previous formulations of the supergravity theories, some of which also retain aspects of the auxiliary fields \cite{Kugo:1982mr,Kugo:1982cu}, were not manifestly covariant under the reparametrizations of these two manifolds. With the methods of \cite{Freedman:2016qnq} we can rewrite all essential formulas in a more geometric fashion, such that reparametrization invariance is manifest. This formulation is very useful in the context of the transition from the superconformal formulation of matter-coupled supergravity to the super-Poincar\'{e} theory. A key point is that the covariant transformations of the superconformal theory maintain  covariance after gauge-fixing to the super-Poincar\'{e} theory.
This allows us to obtain the supersymmetry transformations of the fields in the physical Poincar\'{e} supergravity theory in a straightforward way from those in the superconformal theory.

Though many ingredients of our work apply to both ${\cal N}=2$ and ${\cal N}=1$ theories, we will restrict ourselves in this paper to ${\cal N}=1$ supergravity coupled to chiral multiplets. The Weyl multiplet is the gauge multiplet that allows local supersymmetry. Its vector gauge field $A_\mu $ is an auxiliary field, and we will use its field equations. On the other hand, the auxiliary fields of the chiral multiplets will remain off-shell. We will derive the covariant supersymmetry transformations and the action for the full field content $\{ z^\alpha, \chi^\alpha, h^\alpha \}$ of chiral multiplets. Gauge multiplets can be included in this new formulation, but they will rather appear as spectators in the theory of the chiral multiplets, and we omit them here.

The K\"{a}hler manifolds in ${\cal N}=1$ and ${\cal N}=2$ $D=4$ supergravity are projective manifolds embedded in larger K\"{a}hler manifolds that have conformal properties. The formulation of Poincar\'{e} supergravity as a broken superconformal theory makes use of `compensating fields', whose presence allows the super-Poincar\'{e} group to be promoted to a superconformal group. We start with $n+1$ chiral multiplets, with complex scalar fields $X^I$, $I=0,\ldots ,n$. This includes the compensating scalar field and others scalars that will be physical in the super-Poincar\'{e} theory. However, we do not want to specify which of the fields $X^I$ is the compensator. That is part of the reparametrization invariance that we do not want to break.
We will use the name `embedding manifold' for the scalar manifold with $n+1$ complex fields, and `projective manifold' for the complex $n$-dimensional manifold that describes the super-Poincar\'{e} theory.

We start in section~\ref{ss:geomtransf} by reviewing the main general results needed for a covariant formulation, as found in \cite{Freedman:2016qnq}. These will be exploited in full in this paper.
In section~\ref{ss:sc-off-shell} we discuss the superconformal theory in the covariant formulation, emphasizing the role of the various symmetries. This extends the results of section 3 in \cite{Freedman:2016qnq} to include superconformal transformations.
We present the covariant transformation rules and the covariant form of the action. In section~\ref{ss:sc-projective} we introduce the convenient variables to discuss the super-Poincar\'{e} theory. We discuss the gauge fixing that leads to the Poincar\'{e} group and the resulting projective space. The relation between superconformal and \Poincare supersymmetry is presented in section~\ref{sec:SUSY}. With these preliminaries in place, one can derive the transformation laws of the Poincar\'{e} fields smoothly, given the superconformal transformations. We obtain the full transformation rules, and the part of the action relevant for auxiliary fields,  in section \ref{ss:offshellPoincSymm}. We conclude with a synopsis, and a brief discussion of applications to constrained multiplets in section~\ref{sec:conclusions}.

\section{Geometrisation of transformations}
\label{ss:geomtransf}

We summarize the main ideas and results of \cite{Freedman:2016qnq} for a theory that contains scalar fields $\phi^i(x)$ that are maps from spacetime to coordinate charts on a Riemannian target space $M$ with metric $g_{ij}(\phi)$ and Christoffel connection $\G^i_{jk}$.  We require covariance under reparametrizations
\begin{equation}
  \phi ^i \rightarrow  \phi '^i(\phi )\,.
 \label{phirepar}
\end{equation}
The theory may as well contain composite vectors  $V^i(\phi)$,  such as Killing vectors, and other fields such as the fermions $\chi^i(x)$ of supersymmetric theories, which transform as sections of the tangent bundle of $M$. Their transformation laws\footnote{We use the notation $V^i(\phi )$, to indicate an equation that is only valid when $V^i(\phi) $ is a function of the scalars and not of other fields of the theory.}
\begin{align}
V^i(\phi) \to  V'^i(\phi'(\phi )) = \frac{\pa\phi'^i}{\pa\phi^j}V^j(\phi)\,, \qquad
 \chi ^i \rightarrow \chi^{\prime i}=\frac{\partial \phi^{\prime i}}{\partial \phi ^j}\chi ^j\,,
 \label{psireparameters}
\end{align}
are similar, but there is an important difference that we discuss shortly. As in most treatments of supersymmetry, the fields $\phi ^i$ and $\chi ^i$ are considered as independent, so that $\{\phi ^i,\chi ^i\}$ form a basis of the field space.

Readers are probably familiar with the following definition of covariant spacetime derivatives\footnote{We use torsionless connections.} of $\phi^i,~V^i(\phi),~\chi^i$:
\begin{eqnarray}
 \nabla_\m \phi^i&=&\pa_\m\phi^i\,,\nonumber\\
 \nabla _\mu V^i(\phi )&=&(\partial _\mu \phi ^j) \nabla _j V^i(\phi )= (\pa_\m\phi^j)\left[\pa_j V^i(\phi) +\G^i_{jk}V^k(\phi)\right]\,, \nonumber\\
 \nabla_\m \chi^i &=& \pa_\m\chi^i +\G^i_{jk}\chi^k(\pa_\mu \phi^j)\,.
 \label{covder}
\end{eqnarray}
Using (\ref{phirepar}-\ref{psireparameters}) and the transformation property of $\Gamma ^i_{jk}$, one can show that these covariant derivatives transform as vectors.

Consider an infinitesimal symmetry operation
\begin{equation}
  \delta \phi ^i(\phi ,\chi )\,,\qquad \delta \chi ^i(\phi ,\chi )\,,
 \label{symmetriesphipsi}
\end{equation}
on our system such as spacetime translations or supersymmetry. Then (\ref{phirepar}-\ref{psireparameters}) show that $\d\phi^i$ is a vector, but $\delta \chi^i$ and the induced transformation $\d V^i(\phi )$ are not.  Again we need a connection term to define  \emph{covariant transformations} in the last two cases: \cite{Bergshoeff:2002qk}, \cite[App.14B]{Freedman:2012zz}
\begin{eqnarray}
\hat{\delta }V^i(\phi )&\equiv& \d V^i (\phi)+\G^i_{jk}\d\phi^j V^k(\phi)=\d\phi^j(\pa_jV^i +\G^i_{jk}V^k(\phi)\,, \nonumber\\
\hat\d\chi^i &\equiv& \d\chi^i +\G^i_{jk}\d\phi^j \chi^k\,
\label{covsym}
\end{eqnarray}

We can now observe the difference between the covariant rules for composite vectors $V^i(\phi)$ and vector-valued fields such as $\chi^i(x)$. Only the former can be expressed in terms of covariant derivatives on $M$.
To make this clear we repeat
\begin{equation}
\nabla _\mu V^i(\phi )=(\partial _\mu \phi ^j) \nabla _j V^i(\phi )\,,\qquad \hat{\delta }V^i(\phi )= (\delta  \phi ^j) \nabla _j V^i(\phi )\,.
 \label{nablaonscalarfncts}
\end{equation}
Note that the covariant rules defined above for vectors can be easily extended to covectors and tensors. For the metric of which $\Gamma ^i_{jk}$ are the Christoffel symbols:
\begin{equation}
\nabla_\m g_{ij}(\phi)=(\pa_\m\phi^k) \nabla _k g_{ij}=0\,,\qquad   \hat{\delta }g_{ij}= (\delta  \phi ^k) \nabla _k g_{ij}=0\,.
 \label{hatdeltag0}
\end{equation}

We now state a principle that is both obvious when thought about and  powerful in operation: \emph{
If an action is built as a scalar from vectors and tensors, then invariance under a symmetry operation $\delta $ is equivalent to invariance under the covariant transformation $\hat\d$.}

Ordinary derivatives and transformations commute by definition:
\begin{equation}
  \delta \partial _\mu =\partial _\mu \delta\,,
 \label{deldisddel}
\end{equation}
but the commutator of $\hat\delta $ and $\nabla $ gives rise to curvature terms.
\begin{equation}
\hat{\delta }\nabla _\mu V^i = \nabla _\mu \hat{\delta }V^i+ R_{k\ell}{}^i{}_jV^j(\delta \phi ^k)(\partial _\mu \phi ^\ell)\,.
 \label{commcovderdelta}
\end{equation}
Furthermore, curvature terms appear also in the commutator of covariant derivatives and the commutator of covariant transformations.  This relation as well as those below, which were derived in \cite{Freedman:2016qnq}, are valid both for composite $V^i(\phi)$ and vector-valued fields such as $\chi^i(x)$;
\begin{align}
  &[ \nabla_\mu , \nabla_\nu ] V^k  = R_{ij}{}^k{}_\ell (\partial_\mu \phi^i) (\partial_\nu \phi^j) V^\ell \,,  \nonumber\\
  &   \left[\delta_1,\delta_2\right] V^i=\delta_3 V^i\  \implies\ \left[\hat{\delta}_1,\hat{\delta}_2\right] V^i=\hat{\delta }_3 V^i + R_{k\ell}{}^i{}_j V^j(\delta_1\phi ^k)( \delta  _2\phi ^\ell)
\,,
\label{commcovdelta}
\end{align}
where $\delta_1$ is a shortcut for $\delta[\epsilon _1])$, \ldots  and $\epsilon _3$ is the function of $\epsilon _1$ and $\epsilon _2$ determined by the structure of the symmetry algebra.

We close this section with an exercise for interested readers. {\it  Let $k^i(\phi)$ be a Killing vector on $M$ that acts on fermions fields as    $\d \chi^i = \pa_jk^i\chi^j$. By the rules stated above the covariant form of this symmetry operation is $\hat\d\chi^i =\nabla_jk^i \chi^j$.  Without peeking at \cite{Freedman:2016qnq}, show that
$\hat\delta\nabla_\m \chi^i = \nabla_jk^i\nabla_\m\chi^j$.}

\section{Covariant superconformal theory}
\label{ss:sc-off-shell}

In the first stage of the superconformal  approach to ${\cal N}=1,~D=4$ supergravity,  a set of chiral multiplets, denoted by $\{X^I, ~\Omega^I,~ F^I\}$, with Weyl weight~1, is coupled to the  Weyl multiplet $\{e_\mu ^a,\,\psi _\mu ,\, b_\mu ,\,A_\mu \}$. The complex scalar fields $X^I(x)$ are coordinates of a K\"{a}hler manifold with conformal symmetry. A conformal \Kahler manifold obeys certain homogeneity conditions, which we explain in the next subsection.  These conditions constrain the allowed reparametrizations and also induce a chiral symmetry.
We  apply the geometric methods of section \ref{ss:geomtransf} to define covariant derivatives and covariant transformations,  which transform properly under homogeneous reparametrizations and chiral transformations.  We then focus on covariant superconformal transformations of   $\{X^I, ~\Omega^I,~ F^I\}$ and also write the superconformal action that determines their dynamics. We refer to this setting as the geometric superconformal theory. This prepares the way for a covariant treatment of the physical supergravity theory  in section \ref{ss:sc-projective}.

\subsection{Superconformal K\"{a}hler manifolds}
\label{ss:scKahler}

We first discuss the embedding space spanned by the scalars of the superconformal theory, together with a covariant formulation of its symmetries. In outline,  our discussion  follows  \cite{Freedman:2016qnq}, but we emphasize two new ingredients, namely homogeneity and chiral symmetry, called $T$-symmetry.

The scalars $X^I$ are  coordinates of a \Kahler manifold. Its metric is determined by a \Kahler potential as usual:
\begin{equation}
  G_{I\bar J} = N_{I\bar J} \equiv \partial _I\partial _{\bar J}N(X,\bar X) \,.
 \label{GN}
\end{equation}
In order to apply it to the physical supergravity theory, as we will discuss it in section~\ref{ss:sc-projective}, the metric should have signature  $(-,+,\dots +)$, which corresponds to the index values $I=(0,1, \ldots n$). The negative direction corresponds to the conformal compensator,  but in this fully covariant approach we need not identify it more specifically.

\paragraph{Homogeneity.}  An important condition imposed by superconformal symmetry is that the \Kahler potential $N=N(X,\bar X)$ is homogeneous of weight one\footnote{This means that the manifold possesses a `closed homothetic Killing vector' \cite{Sezgin:1995th} (summarized in\cite[section 15.7]{Freedman:2012zz}). The presence of such a vector $k_{\rm D}{}^I$ implies conformal symmetry, and in a K\"{a}hler manifold, it further implies a Killing vector for the $T$-symmetry: $k_{\rm T}{}^I=\rmi k_{\rm D}{}^I$. In this paper we choose coordinates where this closed homothetic Killing vector is aligned in the direction $k_{\rm D}{}^I = X^I$. A generalization to more general coordinates is possible.} in both the holomorphic and the anti-holomorphic coordinates.
The homogeneity condition requires the equations (in a notation where subscripts on $N$ indicate derivatives)
 \begin{eqnarray}
N(X,\bar X) &=& X^IN_I =\bar X^{\bar I}N_{\bar I} = N_{I\bar J}X^I \bar X^{\bar J}\qquad N_I= N_{I\bar J}\bar X^{\bar J}\,,\nonumber\\
X^IN_{IJ}&=&0\qquad N_{IJ\bar K}\bar X^{\bar K}= N_{I J}\qquad X^KN_{KI\bar J} =0\,.
\label{homogNcons}
\end{eqnarray}
An important consequence of homogeneity is that geometrical quantities, such as the  connection and
curvature tensor\footnote{In general, closed homothetic Killing vectors are zero-modes of the curvature.} for the K\"{a}hler manifold, have zero vectors, viz
\begin{eqnarray}
  \Gamma ^I_{JK}&=& G^{I\bar L}N_{\bar LJK}\,,\qquad X^J\Gamma ^I_{JK}=0\,,
 \label{connectionprop}
\\  R_{I\bar JK\bar L}&=& N_{I\bar JK\bar L}-N_{IK\bar M}G^{\bar MN}N_{N\bar J\bar L}\,,\qquad X^I R_{I\bar JK\bar L}=0\,.
 \label{zerovectorR}
\end{eqnarray}
K\"{a}hler transformations of $N(X,\bar X)$ are not permitted since holomorphic additional terms do not satisfy the homogeneity requirement.

\paragraph{Chiral Symmetry.}  Homogeneity of the superconformal \Kahler manifold,  together with its complex structure,  imply that there are separate dilatation and  chiral symmetries under which the scalars $X^I,\, \bar X^{\bar J}$ transform as
\begin{equation}
  \d X^I = (\l_D + \rmi \l_T) X^I\,,\qquad  \d \bar X^{\bar I} = (\l_D - \rmi \l_T) \bar X^{\bar I} \,.
 \label{chiraltransfoX}
\end{equation}
These are the Weyl scaling and the chiral $T$-symmetry, which is the $\U(1)$ R-symmetry of the conformal supersymmetry algebra. The Lagrangian contains auxiliary connections for both symmetries. The $T$-connection is the  gauge field $A_\mu(x)$ of the Weyl multiplet.\footnote{It is called ${\cal A}_\mu $ in this section in which we include only its action on the scalars.}
We  focus on the $T$-connection, since the dilatation gauge field $b_\mu$ will be set to zero when gauge fixing the special conformal transformations in the passage to the \Poincare theory. The scalar Lagrangian is therefore
\be \label{Lemb}
L = - N_{I\bar J} g^{\m\n} \nabla_\m X^I \nabla_\nu \bar X^{\bar J} \,,
\ee
where
\begin{equation}
  \nabla _\mu X^I= \partial _\mu X^I- \rmi{\cal A}_\mu X^I\,.
 \label{nablaX}
\end{equation}
and the T-connection $\ca_\m$ transforms as  $\d\ca_\m = \pa_\m \l_T$.
The connection is an auxiliary field whose field equation is solved by
\be\label{Aconn}
\rmi\ca_\m = \frac{1}{2N}\left(N_I\pa_\m X^I-N_{\bar J}\pa_\m\bar X^{\bar J} \right)\,.
\ee
After substitution of this result in (\ref{Lemb}), we find the equivalent Lagrangian
\be \label{Lemb2}
L = -\frac{1}{4N} \pa_\m N\pa^\m N - N (\pa_\m X^I) (\pa^\m\bar X^{\bar J})\partial _I\partial _{\bar J} \ln N\,.
\ee
Upon redefining $N = - r^2$ this can be interpreted as a cone over a projective manifold. Note that the radial direction has a kinetic term of the wrong sign. However, this corresponds to the conformal compensator and not to a physical field and does not pose a problem.

The composite connection ${\cal A}_\mu$ in (\ref{Aconn}) must be included in our covariant definitions as follows.
A general vector $V^I$ of chiral weight $c$ satisfies
\begin{align}
   \delta _T V^I=\rmi c\lambda _T V^I \,,
  \label{Ttrans}
 \end{align}
and we \emph{extend} the definition of covariant derivatives $\nabla_\mu$, (\ref{covariantdef}), with the $T$-connection:
\be\label{cDhatX}
\nabla _\mu V^I \equiv (\pa_\m -\rmi c {\cal A}_\m)V^I +\G^I_{JK} V^J\partial _\mu  X^K\,.
\ee
This transforms as a tangent vector under coordinate reparametrizations, and has chiral weight $c$.  Similarly, we define covariant transformation rules as
\begin{equation}
  \hat{\delta }V^I \equiv \delta V^I +\Gamma ^I_{JK}V^J\delta X^K \,,
 \label{hatdeltaVI}
\end{equation}
which also transform as a tangent vector.

In  the special case of composite vectors on the target space, i.e. $V^I = V^I(X,\bar X)$, the $T$-transformation is implemented as the Killing symmetry
\begin{align}
 \delta _T V^I(X,\bar X) = \rmi \lambda _T(X^J\partial _J-\bar X^{\bar J}\partial _{\bar J})V^I(X,\bar X)\,.
 \label{Ttransfsplit}
 \end{align}
Covariant spacetime derivatives and transformation rules satisfy relations to covariant derivatives on the K\"{a}hler manifold, i.e.
\begin{equation}
  \nabla_J V^I= \partial_J V^I +\Gamma ^I_{JK}V^K\,,\qquad  \nabla_{\bar J} V^I= \partial_{\bar J} V^I\,,
 \label{nablaJVI}
\end{equation}
which generalize (\ref{nablaonscalarfncts}),
\begin{equation}
  \nabla _\mu V^I(X,\bar X)= \nabla _\mu X^J \, \nabla_J V^I + \nabla _\mu \bar X^{\bar J} \, \nabla_{\bar J} V^I \,,  \qquad  \hat{\delta }V^I(X,\bar X)  = \delta X^J\nabla _J V^I+ \delta X^{\bar J}\overline{\nabla}  _{\bar J} V^I \,,
 \label{nablaconfscalar}
\end{equation}
since the $T$-connections follow this pattern on account of (\ref{Ttransfsplit}).
The last equation is valid separately for all transformations, e.g. both for supersymmetry and for $T$-transformations.
Note also that  (\ref{nablaX}) is consistent with (\ref{cDhatX}) because of  (\ref{connectionprop}), and that
$\hat{\delta }X^I =\delta X^I$.

In the rest of this paper we develop covariant formulas under homogeneous reparametrizations\footnote{If we use results of actions and transformations in a more general frame than frames where $k_{\rm D}^I=X^I$, the restriction to homogeneity  can be removed.} of the target space.  This means that coordinates transform as vectors under this class of reparametrizations,
\begin{align}
  X^I \ \rightarrow \ X'^I (X)= X^I \partial_I X'^I(X) \,,
 \label{homogpreservX}
\end{align}
which is not true for the more general coordinate transformations of section \ref{ss:geomtransf}.
\subsection{Covariant superconformal transformations}
\label{ss:susytransfcov}

In this section we obtain the covariant form of the \susy variations given in \cite[(17.3)]{Freedman:2012zz}. The rules for target space covariance are essentially as given in section 3 of \cite{Freedman:2016qnq}, but we extend them to incorporate two features of superconformal supergravity.  First we have local $S$-supersymmetry with parameter $\eta $. Second, the superconformal covariant derivatives include connections of the Weyl multiplet.  The covariant transformations are\footnote{We use the notation that $\Omega ^I$ is left-handed, i.e. $\Omega ^I=P_L\Omega ^I$ and
$\Omega ^{\bar I}=P_R\Omega ^{\bar I}$.}
\begin{align}
  \hat{\delta }X^I&=\delta X^I = \frac{1}{\sqrt{2}} \bar \epsilon \Omega^I\,,
  \nonumber\\
  \hat{\delta }\Omega ^I &=\frac{1}{\sqrt2} P_L\left({\slashed{\cd }}X^I + \hat{F}^I\right) \epsilon  +\sqrt2 X^I P_L\eta\,,\nonumber\\
\hat{\delta }\hat F^I&=\frac{1}{\sqrt{2}}\bar\epsilon\left[ \gamma^\mu\hat{\cal D}_\mu\Omega ^I+\ft12R_{J\bar L}{}^I{}_K \Omega ^{\bar L}\overline{\Omega}{} ^J\Omega^K\right]\,,
\label{confcovtransf}
\end{align}
where we introduced the covariant auxiliary field
\begin{equation} \label{covariantF}
\hat F^I = F^I - \frac12 \Gamma^I_{JK}\bar \Omega^J\Omega^K\,.
\end{equation}
The superconformal covariant derivatives depend on the fields of the Weyl multiplet: the frame field $e_\mu ^a$, the gravitino $\psi_\mu $, the dilatation gauge field $b_\mu $ and the $T$-gauge field $A_\mu $. These are independent fields. After the action is constructed, the field equation of $A_\mu $ sets it equal to ${\cal A}_\mu $ as in (\ref{Aconn}) plus a fermionic part given in (\ref{Amusplit}) below. These derivatives are
\begin{align}
  \cd_\mu  X^I =&  \partial  _\mu X^I-b_\mu X^I -\rmi A_\mu  X^I -\frac{1}{\sqrt{2}}\bar \psi _\mu \Omega ^I\,,  \label{covderconf}\\
  {\cal D}_\mu\Omega ^I =& \left(\partial  _\mu+\frac{1}{4}\omega _\mu {}^{ab}\gamma _{ab}-\frac{3}{2}b_\mu +\frac{1}{2}\rmi A_\mu \right)\Omega ^I-\frac{1}{\sqrt{2}}P_L\left(\slashed{\cal D}X^I+ F^I\right)\psi _\mu -\sqrt{2}P_LX^I\phi _\mu \,,
\nonumber
\end{align}
where $\phi _\mu$ is the gauge field of the $S$-supersymmetry. This is dependent on the other fields
\begin{equation}
  \phi _\mu  = - \ft12\gamma ^\nu R'_{\mu \nu }(Q)+\ft{1}{12}\gamma _\mu \gamma
  ^{\rho \nu }R'_{\rho \nu }(Q)\,,\qquad R'_{\mu \nu }(Q)=2
   \left( \partial _{[\mu }+\ft14\omega _{[\mu }{}^{ab}\gamma _{ab}+\ft12b_{[\mu }-\ft32\rmi
A_{[\mu }\gamma _*\right) \psi
_{\nu ]}\,.
 \label{valuephimu}
\end{equation}
The spin connection used here is the conformal connection with gravitino and $b_\mu $ torsion
\begin{equation}
  \omega _\mu {}^{ab}= 2 e^{\nu[a} \partial_{[\mu} e_{\nu]}{}^{b]} -
e^{\nu[a}e^{b]\sigma} e_{\mu c} \partial_\nu e_\sigma{}^c+2e_\mu {}^{[a}e^{b]\nu }b_\nu
+\ft12\bar \psi
  _\mu \gamma ^{[a}\psi ^{b]}+\ft14\bar \psi ^a\gamma _\mu \psi ^b\,.
 \label{solomebpsi}
\end{equation}
The geometric covariant derivative is\footnote{Since $Z^\alpha $ of \cite{Freedman:2016qnq} was a coordinate and not a vector, we did not define a covariant derivative $\nabla _\mu Z^\alpha $. Here $X^I$ is also a vector, and thus in principle we can define the geometric covariant $\wcD_\mu  X^I$. However, due to (\ref{connectionprop}), this is equal to $\cd_\mu X^I$.}
\begin{equation}
  \wcD_\mu \Omega ^I= {\cal D}_\mu \Omega ^I +\Gamma ^I_{JK}\Omega ^J{\cal D}_\mu X^K\,.
 \label{wcD}
\end{equation}

The calculations to arrive at (\ref{confcovtransf}) are the same as in section 3 of \cite{Freedman:2016qnq}. We only have the extra $S$-supersymmetry.
The covariance of the $S$ transformation part\footnote{In a frame with an arbitrary closed homothetic Killing vector, the $S$-supersymmetry transformations of $\Omega ^I$ would be of the form
$  \delta _S \Omega ^I = \sqrt{2} k_{\rm D}{}^IP_L\eta$, and $k_{\rm D}{}^I$ transforms to other frames as a vector.} of $\Omega ^I$, is the statement that $X^I$ behaves as a vector. This is only true for transformations of the form (\ref{homogpreservX}).

There is no $S$-transformation of $\hat{F}^I$ in (\ref{confcovtransf}), because $F^I$ does not transform and the $S$-transformations of the fermions in (\ref{covariantF}) do not contribute due to (\ref{connectionprop}).\footnote{If we would formulate the transformations with an arbitrary closed homothetic Killing vector, $F^I$ would transform under $S$, and the term in this extra part of $\hat{F}^I$, which is then $\G^I_{JK}k_{\rm D}{}^K$, would cancel the $S$-transformations of $F$, such that $\hat{F}^I$ would still be $S$-invariant.}

The commutator relation applied on $\Omega ^I$ is similar to  (\ref{commcovdelta})
\begin{eqnarray}
 \left[\hat{\delta} _1,\hat{\delta}_2\right]\Omega ^I&=&\hat{\delta }_3\Omega ^I + R_{K\bar L}{}^I{}_J\Omega ^J\left[\ft12\bar \epsilon _1\Omega ^K\,\bar \epsilon _2\Omega ^{\bar L}- (1\leftrightarrow 2))\right]\nonumber\\
  & = & \hat{\delta }_3\Omega ^I -\ft14 R_{K\bar L}{}^I{}_J\left[\epsilon _1(\bar \Omega ^J  \Omega ^K)\bar \epsilon _2\Omega ^{\bar L} - (1\leftrightarrow 2))\right]
 \label{explicitcommOmega}
\end{eqnarray}
Since $X^I$ can be viewed as a vector, we might expect that  $\left[\hat{\delta}_1,\,\hat{\delta}_2\right]X ^I$ should include an analogous curvature term.  But it  vanishes due to (\ref{zerovectorR}).

\subsection{Superconformal action}
\label{ss:covaction}
There are two independent  parts of the superconformal action for chiral multiplets. The first is the covariant kinetic action, called $[N ]_D$ because it is a $D$-term which requires the K\"{a}hler potential as input data.
The second part is the superpotential action, an $F$-term called $ [\mathcal{W}]_F$, which is determined by the holomorphic superpotential ${\cal W}(X)$.  Each part is invariant under the transformation rules (\ref{confcovtransf}).

The kinetic action of \cite[(17.19)]{Freedman:2012zz} simplifies considerably when one uses covariant derivatives and $\hat F^I$. In this geometric formulation, it can be written as (it actually includes kinetic terms for graviton and gravitino in the last line)
\begin{eqnarray}
 \lefteqn{  [N ]_D\,e^{-1}=
N_{I\bar J}\left( -{\cal D} _\mu X^I {\cal D}^\mu \bar X^{\bar J}-\frac12\bar \Omega^I
  {\slashed{\wcD }}\Omega^{\bar J}-\frac12\bar \Omega^{\bar J}
\slashed{\wcD}\Omega^I
+\hat F^I \bar {{\hat F}}^{\bar J }
\right)}
\nonumber\\
&&\quad +\,\frac14R_{I \bar K J \bar L}\,\bar \Omega^I
 \Omega^J \bar \Omega^{\bar K}\Omega^{\bar
L}+\left[ \frac1{2\sqrt{2}}\bar \psi \cdot \gamma
\left(N_{I\bar J}\hat F^I\Omega^{\bar J}-N_{I\bar J}\slashed{\cal D}\bar X^{\bar J}\Omega^I
\right) \right.
\nonumber\\
&&\quad \left.+\,\frac{1}{8}\rmi\varepsilon ^{\mu \nu \rho \sigma }\bar \psi _\mu \gamma _\nu \psi _\rho
\left( N_I{\cal D}_\sigma X^I+\frac{1}{2}N_{I\bar J}\bar \Omega^I\gamma _\sigma \Omega^{\bar J}
+\frac{1}{\sqrt{2}}N_I\bar \psi _\sigma \Omega^I\right)
+\hc\right]\nonumber\\
&&\quad +\,\frac16N\left(-R(\omega )
+\ft12\bar \psi _\mu \gamma ^{\mu \nu \rho } R'_{\nu \rho }(Q)
\right) -\frac{1}{6\sqrt{2}}\left( N_I \bar \Omega^I+N_{\bar I}\bar \Omega^{\bar I}\right) \gamma ^{\mu \nu } R'_{\mu
\nu }(Q)\,.
 \label{kinchiralConf}
\end{eqnarray}

The superpotential  is a homogeneous, holomorphic function of Weyl weight 3:
\begin{equation}
X^I\mathcal{W}_I = 3 \mathcal{W},\qquad\qquad \mathcal{W}_I \equiv \frac{\partial}{\partial X^I}\mathcal{W}.
\end{equation}
The $F$-term action is given by
\begin{equation}
  [\mathcal{W}]_Fe^{-1}=
{\cal W}_I \hat F^I -\ft12{\cal  W}_{I;J}\bar \Omega^{I}\Omega ^{J}
 +\frac{1}{\sqrt{2}}{\cal  W}_I\bar \psi\cdot
\gamma  \Omega^{I}+\ft12 {\cal  W}\bar \psi _{\mu }P_R \gamma ^{\mu \nu
}\psi _{\nu }+ \hc\,,
 \label{WF}
\end{equation}
where the semicolon sign is used for covariant derivatives, e.g. ${\cal  W}_{I;J}={\cal  W}_{IJ}-\Gamma _{IJ}^K{\cal W}_K$. The Ricci scalar $R(\omega)$ is calculated with the spin connection in (\ref{solomebpsi}), and $R'_{\mu \nu }(Q)$ is given in  (\ref{valuephimu}).

\subsection{On shell transformation of auxiliary fields}

The auxiliary field $\hat{F}^I$ can be eliminated using the algebraic field equation\footnote{Note that $\hat{F}^I=0$ on shell is valid both in the rigid limit and in supergravity when ${\cal W}=0$.
The reason is that the $F\g\cdot \psi$ term on the 2nd line of (\ref{kinchiralConf}) cancels with a similar term in
the supercovariant derivative of $\Omega^I$, see (\ref{covderconf}), in the fermion kinetic term.}
 \begin{align}
   \hat F^I_{\text on-shell} = -  \overline{\mathcal{W}}_{\bar K} G^{I\bar K} =-\overline{\mathcal{W}}^I \,.
   \label{fehatFI}
 \end{align}
The covariant  transformation of the right side gives
\be
\hat\delta(\hat F^I_{\text on-shell}) =-\frac{1}{\sqrt2} \overline{\mathcal{W}}^I{}_{;\bar J}\, \bar\eps\, \bar{\O}^{\bar J}\,.
\label{delhatFonshell}
\ee
We now show that  this result is consistent with (\ref{confcovtransf}) when the fermion equation of motion is used.   To check this,
it is useful to rewrite the transformation of the auxiliary field as
\begin{align}\label{delFcovDiederik}
\hat{\delta }\hat F^I= & \frac{1}{\sqrt{2}}\bar\epsilon\left[ \gamma^\mu\hat{\cal D}_\mu\Omega ^I+\ft12R_{J\bar L}{}^I{}_K \Omega ^{\bar L}\overline{\Omega} ^J\Omega^K
+ \overline{\mathcal{W}}^I{}_{;\bar J} \Omega^{\bar J} + \gamma \cdot \psi \left( \overline{\mathcal{W}}^{I} + \hat F^I\right)  \right]  \notag \\
 & - \frac{1}{\sqrt{2}} \bar \epsilon  \left[  \gamma \cdot \psi \left( \overline{\mathcal{W}}^{I} + \hat F^I\right) + \overline{\mathcal{W}}^I{}_{;\bar J} \Omega^{\bar J}\right]
 \approx - \frac{1}{\sqrt{2}}\overline{\mathcal{W}}^I{}_{;\bar J} \,\bar \epsilon\,\Omega^{\bar J}\,.
\end{align}
The first line of this equation is equal to the field equation for the fermionic fields. Further, the round brackets in the second line enclose the field equation for the auxiliary fields. The net result is indeed the on-shell transformation (\ref{delhatFonshell})!

\section{Projective space and gauge fixing}
\label{ss:sc-projective}

\subsection{Introducing Poincar\'{e} coordinates}\label{ss:poincoords}

The next important step is to move toward physical variables by the substitution
\begin{equation}
  X^I = y Z^I(z^\a)\,.
 \label{XyZ}
\end{equation}
This relates the $n+1$ complex fields $X^I$  to new variables $z^\a$ with $\a=1,\dots n$ and $y$.  The $z^\a$ are the physical scalars, which are coordinates on an $n$-dimensional projective K\"{a}hler manifold.   Because we require invariance under the reparametrization $z^\a\to z'^{\a}(z)$, the $Z^I(z)$ are arbitrary functions, but subject to the requirement that
\bea \label{ZpaZ}
\begin{pmatrix}
Z^I \cr \pa _{\a} Z^I
\end{pmatrix}
\eea
is a non-singular $(n+1)\times(n+1)$ matrix.

When the \Poincare coordinate Ansatz is substituted in (\ref{Lemb2}), we find that $\pa_\m y$ cancels in the second term (see \cite[Sec 17.3.4]{Freedman:2012zz}),
and the Lagrangian takes the form
\be \label{Lemb3}
L = -\frac{1}{4N} \pa_\m N\pa^\m N - N (\pa_\m z^\a )(\pa^\m \bar z^{\bar \b})\pa_\a\pa_{\bar \b}\ln\left[Z^IG_{I\bar J}\bar Z^{\bar J}\right]\,,\qquad N=y\bar y Z^IG_{I\bar J}\bar Z^{\bar J}\,.
\ee
Interpreting this as a cone, the projective manifold is a non-linear $\sigma$-model whose metric is given by the K\"{a}hler potential
\begin{equation}
  {\cal K}(z,\bar z)= - a\ln \left[-a^{-1}\, Z^I(z)G_{I\bar J}\bar Z^{\bar J}(\bar z)\right] = a\ln \left[ - \frac{a y \bar y}{N} \right]\,,
 \label{defcalK}
\end{equation}
where we have included an arbitrary constant $a$ for later convenience. Moreover, in these coordinates, the chiral connection (\ref{Aconn}) takes the form
(see \cite[Ex.17.10]{Freedman:2012zz})
\begin{equation}
  \rmi\,{\cal A}_\mu = \ft12a^{-1}\left( \partial _\mu \bar z^{\bar \alpha }\partial _{\bar \alpha }{\cal K}
  - \partial _\mu z^\alpha \partial _\alpha {\cal K}\right)+\ft12 \partial_\mu \ln(y/\bar{y}) \,.
\label{valueAmuproj}
\end{equation}
Since this Lagrangian is simply a reparametrization of  (\ref{Lemb}), it inherits its chiral and dilatation symmetries.

However, there is  now a new symmetry since the split of coordinates (\ref{XyZ}) is not unique: it is invariant under
\be\label{Ktrfs}
  Z^I\to \rme^{-f(z)/a}Z^I\,,\qquad y\to \rme^{f(z)/a}y\,.
 \ee
This induces K\"{a}hler transformations on the potential:
\begin{equation}
  {\cal K}(z,\bar z)\to  {\cal K}(z,\bar z) +f(z) + \bar f(\bar z)\,.
 \label{delcalKf}
\end{equation}
We consider the transformations with $f$ and $\bar f$ as independent.
Since we have these transformations we can choose the dilatations and $T$-transformation to act only on $y$ and not on $Z^I$;
\begin{equation}
  \d_{\rm D} y = \l_{\rm D} y\,,\qquad \d_T y = \rmi\l_T y\,,\qquad \delta _{\rm D}z^\alpha =\delta _Tz^\alpha =0\,.
 \label{DandTony}
\end{equation}
Hence $y$ and $Z^I$ have chiral weights $1$ and $0$, respectively. More generally, functions of the scalars $y$ and $z^\alpha$ can have  weights $c$, $w_+$ and $w_-$ and transform as
\begin{equation}
  \left(\delta _T[\lambda _T]+\delta _{\rm K}[f]\right)V=\left( \rmi c \lambda _T-a^{-1}w_+ f(z) -a^{-1}w_- \bar f(\bar z) \right)V\,,
 \label{defcw}
\end{equation}
under $T$ and K\"{a}hler transformations.

In order to define covariant derivatives, we introduce auxiliary connections for K\"{a}hler transformations\footnote{Note that in \cite[(17.162)]{Freedman:2012zz} we used another normalization for $\omega _\alpha $ and $\omega _{\bar \alpha }$, because we introduced them as gauge fields of the symmetry $\Im f$. Here we consider $f$ and $\bar f$ as independent transformations, where the first one has only holomorphic gauge connection as in the equation below, and the second one has only anti-holomorphic gauge connection.}
\begin{equation}
\o_\m = \o_\a\pa_\m z^\a\,,\qquad  \omega _\alpha =\partial _\alpha{\cal K}\,,\qquad \bar  \o_\m = \bar \o_{\bar \alpha }\pa_\m \bar z^{\bar \alpha }\,,\qquad \bar \omega _{\bar \alpha} =\partial _{\bar \alpha}{\cal K}\,.
 \label{omaldef}
\end{equation}
The covariant derivatives are then defined as
\begin{equation}
  \nabla _\mu V \equiv \left(\partial _\mu -  \rmi c {\cal A}_\mu +a^{-1}w_+ \omega _\mu  +a^{-1}w_- \bar \omega _\mu  \right)V\,.
 \label{defnabmuV}
\end{equation}
For tensors there are the extra terms such as the last term in (\ref{cDhatX}). The weights for various fields and quantities are given in the left part of Table \ref{tab:weights}.
\begin{table}[t]
\[
  \begin{array}{|c||ccc||cc|}
    \hline
    & c & w_+ & w_- & {\hat w}_+ & {\hat w}_-\\  \hline \hline
    &&&&&\\[-4mm]
     X & 1 & 0 & 0 & \nicefrac12 & - \nicefrac12 \\
     y & 1 & -1 & 0 & - \nicefrac12 & - \nicefrac12 \\
     Z^I  & 0 & 1 & 0 & 1 & 0 \\
 z^\alpha & 0 & 0 & 0 & 0 & 0 \\
\hline
&&&&&\\[-4mm]
      \Omega ^I & -\nicefrac12 & 0 & 0 & - \nicefrac14 & \nicefrac14 \\
     \chi^0, \chi ^{\alpha} & -\nicefrac32 & 0 & 0 & - \nicefrac34 & \nicefrac34 \\
\hline
&&&&&\\[-4mm]
   F^I & -2 & 0 & 0 & -1 & 1 \\
  h^0 , h^\alpha & -3 & 0 & 0 & - \nicefrac32 & \nicefrac32  \\
   \hline
   &&&&&\\[-4mm]
    {\cal W} & 3 & 0 & 0 & \nicefrac32 & - \nicefrac32  \\
     W & 0 & 3 & 0 & 3 & 0 \\ \hline
   \end{array}
\]
\caption{\it The chiral and K\"{a}hler weights of superconformal and super-Poincar\'{e} fields. The weights of the complex conjugate fields are obtained by $c\leftrightarrow -c$, $w_+\leftrightarrow w_-$ and ${\hat w}_+\leftrightarrow{\hat w}_-$. }
\label{tab:weights}
\end{table}

In particular, the covariant derivative $\nabla _\mu y$ is thus
\begin{equation}
 \nabla _\mu  y = \left(\partial _\mu - \rmi{\cal A}_\mu  -a^{-1}\omega _\mu\right) y =  \frac12 y \partial _\mu (\ln y\bar y)-\frac{1}{2a}y\partial _\mu {\cal K}\,,
\label{nablamuyZAfilled}
\end{equation}
while
\begin{equation}
  \nabla _\mu Z^I = \left(\partial _\mu  +a^{-1}\omega _\mu\right)Z^I\,.
 \label{nablamuZI}
\end{equation}
The latter two equations show that these covariant derivatives are an extension of (\ref{nablaX}), consistent with (\ref{XyZ}).

\subsection{Gauge fixing}

Superconformal symmetries that are not part of the Poincar\'e superalgebra must be gauge fixed so as to maintain covariance under target space reparametrization. In this section we discuss the
fixing of the bosonic symmetries, namely dilatation, chiral, and special conformal\footnote{Special conformal transformations are fixed by eliminating one field from the Weyl multiplet: the gauge field of dilations.}
symmetries.

Dilatations are fixed by requiring that $N$ is constant, i.e.
\be
N(X,\bar X) =-a\,.
\label{dilatationgauge}
\ee
For ${\cal N}=1$ supergravity, the value $a=3\kappa ^{-2}$ canonically normalizes the Einstein--Hilbert term. This translates into the following condition on the magnitude of $y$:
\be  \label{yybK}
y\bar y = -a\left[Z^I G_{I\bar J}\bar Z^{\bar J}\right]^{-1}\quad {\rm or}\quad  a\ln (y\bar y)= \ck(z,\bar z)\,.
\ee
With this gauge choice,  (\ref{nablamuyZAfilled}) implies
\begin{equation}
  \nabla_\mu y =0\,.
 \label{nablamuy0}
\end{equation}

To define all variables in terms of $z$ and $\bar z$, we must determine both the modulus of $y(z,\bar z)$ as in (\ref{yybK}), and its phase. This is done by the  chiral symmetry gauge fixing condition
\begin{equation}
  y=\bar y= \rme^{{\cal K}/2a}\,.
 \label{gfT}
\end{equation}
As required, this expresses the field $y$ in terms of the \Poincare fields $(z, \bar z)$.

An important consequence of this gauge choice is that the auxiliary chiral connection (\ref{valueAmuproj}) can now be expressed as the pullback of covariant derivatives on the projective manifold.  For this purpose we write
\begin{eqnarray}
  &&{\cal A}_\mu  = {\cal A}_\alpha \partial _\mu z^\alpha +{\cal A}_{\bar \alpha }\partial  _\mu \bar z^{\bar \alpha }\,,\qquad {\cal A}_\alpha = \rmi\partial _\alpha \ln y=
  \frac{1}{2a}\rmi\partial _\alpha {\cal K}\,.
 \label{calAmualpha}
\end{eqnarray}
The last equation follows from (\ref{yybK}).

\subsection{Covariant derivatives in projective space}
\label{ss:covderproj}
The coordinates of the projective manifold are $z^\alpha $, $\alpha =1,\ldots ,n$ and their complex conjugates $\bar z^{\bar \alpha }$. This manifold has a K\"{a}hler potential ${\cal K}(z,\bar z)$, and corresponding metric
\begin{equation}
  g_{\alpha \bar \beta }=\partial _\alpha \partial _{\bar \beta }{\cal K}(z,\bar z)\,.
 \label{projKametric}
\end{equation}
Functions and tensors on the projective space depend on the spacetime points via their dependence on $z$ and $\bar z$. Therefore we define the split covariant derivatives as
\begin{equation}
  \nabla _\mu V(z,\bar z)\equiv  \nabla _\alpha V \,\partial _\mu z^\alpha + \overline\nabla _{\bar \alpha} V\partial _\mu\bar z^{\bar \alpha }\,,
 \label{nablaVsplit}
\end{equation}
for any scalar quantity $V(z,\bar z)$.

After the gauge fixing, ${\cal A}_\alpha $ and $\omega _\alpha $ are both related to $\partial _\alpha {\cal K}$. As a consequence, only specific combinations of the weights $c$ and $w_\pm$ occur  in the covariant derivatives:
\begin{equation}
  \hat{w }_+ =w_++\ft12c\,,\qquad \hat{w }_-= w_--\ft12 c\,.
 \label{hatom+-}
\end{equation}
These combined weights are also given in Table \ref{tab:weights}. We then have
\begin{equation}
  \nabla _\alpha V =\partial _\alpha V +\frac1a\hat{w }_+V(\partial _\alpha {\cal K})\,,\qquad
  \overline\nabla _{\bar \alpha} V =\partial _{\bar \alpha} V +\frac1a\hat{w }_-V(\partial _{\bar \alpha} {\cal K})\,.
 \label{defnablaV}
\end{equation}
The weights of $\nabla _\alpha V$ are equal to those of $V$.

For quantities that transform as vectors and  tensors under reparametrizations of the embedding or projective space we include the appropriate Levi-Civita connections.  Thus for a vector $V^I$, this becomes
\begin{eqnarray}
  \nabla _\alpha V^I &=&\partial _\alpha V^I+\frac1a\hat{w }_+V^I(\partial _\alpha {\cal K})+\Gamma ^I_{JK}V^J\partial  _\alpha X^K\,,\nonumber\\
  \overline\nabla _{\bar \alpha} V^{\bar I} &=&\partial _{\bar \alpha} V^{\bar I} +\frac1a\hat{w }_-V^I(\partial _{\bar \alpha} {\cal K})+\overline{\Gamma} ^{\bar I}_{\bar J\bar K}V^{\bar J}\partial _{\bar\alpha} X^{\bar K}\,.
 \label{defnablaVI}
\end{eqnarray}
There are no Levi-Civita connection terms for $\nabla _\alpha V^{\bar I}$, but they might have K\"{a}hler connections. The covariant chain rule also holds:
\begin{equation}
  \nabla _\alpha V^I (z,\bar z)= \nabla _\alpha X^J \nabla _J V^I\,.
 \label{nabchainrule}
\end{equation}
Similarly for tensors in the projective space, like $V^\alpha$, we define
\begin{equation}
  \nabla _\alpha V^\gamma = \partial _\alpha V^\gamma + \frac1a\hat{w }_+V^\gamma (\partial _\alpha {\cal K})+\Gamma ^\gamma _{\alpha \beta }V^\beta \,.
 \label{nablaaVbeta}
\end{equation}

As a particular example of these covariant expressions, it follows from \eqref{nablamuy0} that
\be
\nabla_\a y =\overline\nabla _{\bar\a} y =0\,.
\label{nablaay0}
\ee
Hence $y$ is covariantly constant on the projective manifold and in spacetime. This indicates that it is not a physical variable, but it is needed to enforce the Weyl scaling property of (\ref{XyZ}). Similarly, we have
\begin{equation}
    \nabla _\alpha Z^I \equiv \left(\partial _\alpha +a^{-1}(\partial _\alpha{\cal K})\right)Z^I\,, \qquad \overline\nabla _{\bar \a} Z^I \equiv \pa_{\bar \a} Z^I =0\,.
 \label{nablaalZ}
\end{equation}
These equations are often used in the form
\begin{equation}
  \nabla _\alpha X^I = y\, \nabla _\alpha Z^I\,.
 \label{nabXZ}
\end{equation}

Finally, these covariant derivatives can be applied to the metrics in embedding and projective spaces,
\begin{equation}
  \nabla_\a G_{I\bar J}=0\,,\qquad \nabla_\g g_{\alpha \bar \beta }=0\,,
 \label{covdermetric0}
\end{equation}
and hence are compatible with the metrics of these spaces.

For scalar-dependent quantities that are defined in the embedding space, we can relate $\nabla _\alpha $ to the $\nabla _I$ derivatives. In this case, we thus consider quantities $V^I$ built from the $X$ and $\bar X$, which are therefore K\"{a}hler-invariant and $\hat w_\pm = \pm c/2$. Then one can prove that
\begin{equation}
  \nabla _\alpha V^I = \nabla _\alpha X^J \nabla _J V^I\,.
 \label{covchainnabla}
\end{equation}
To prove this one uses the equality of  (\ref{Ttrans}) and  (\ref{Ttransfsplit}) and $\nabla _\alpha \bar X^{\bar I}=0$.

\subsection{Geometric identities}

The general structure can be used to derive a number of geometric identities for these derivatives. First of all, the commutators of covariant derivatives are determined by the curvatures of the different symmetries. There are no curvatures in the commutators of two holomorphic derivatives, but we have
\begin{eqnarray}
 \left[\nabla _\alpha ,\overline{\nabla }_{\bar \beta }\right] V  & = & a^{-1}(-\hat{w}_++\hat{w}_-)g_{\alpha \bar \beta }V\,, \nonumber\\
 \left[\nabla _\alpha ,\overline{\nabla }_{\bar \beta }\right] V^I  & = & a^{-1}(-\hat{w}_++\hat{w}_-)g_{\alpha \bar \beta }V^I + R_{K\bar L}{}^I{}_JV^J (\partial _\alpha X^K)(\partial _{\bar \alpha }\bar X^{\bar L})\,, \nonumber\\
\left[\nabla _\alpha ,\overline{\nabla }_{\bar \beta }\right] V_\gamma   & = & a^{-1}(-\hat{w}_++\hat{w}_-)g_{\alpha \bar \beta }V_\gamma + R_{\alpha \bar\beta \gamma }{}^\delta V_\delta \,.
 \label{commNablagen}
\end{eqnarray}
Note that in the middle line of (\ref{commNablagen}) we could replace $\partial _\alpha X^K$ by $\nabla _\alpha X^K$ due to the curvature properties (\ref{zerovectorR}), and obtain a covariant expression.

The dilatational gauge fixing condition can be written as
\begin{equation}
  y\bar y\, Z^I G_{I\bar J}\bar Z^{\bar J}= -a\,.
 \label{gfZGZ}
\end{equation}
Applying the covariant derivative $\nabla _\alpha $ using (\ref{nablaay0}) and the conjugate of (\ref{nablaalZ}) gives
\be\label{orthog}
y \bar y\,G_{I\bar J}\bar Z^{\bar J}\nabla_\a Z^I=0\,.
\ee
Applying $\overline\nabla _{\bar \b}$ on the latter we obtain
\be \label{Gg}
y\bar y\bar G_{I\bar J}\overline\nabla _{\bar \b}\bar Z^{\bar J}\nabla_\a Z^I = - y\bar y G_{I\bar J}\bar Z^{\bar J}( \overline\nabla _{\bar \b}\nabla_\a Z^I)=  g_{\a\bar \b}\,,
\ee
where we have used the second line of the commutators (\ref{commNablagen}), which implies
 \begin{align}
  [\nabla_\a,\overline\nabla _{\bar \b}] \,Z^I = -  a^{-1}g_{\a\bar\b}\,Z^I\ \implies \  \overline\nabla _{\bar\b}\nabla_\a Z^I =  a^{-1}g_{\a\bar\b}Z^I\,.
\label{commnabZ}
 \end{align}
We write the equations (\ref{gfZGZ})--(\ref{Gg}) collectively as
\begin{equation}
y\bar y \begin{pmatrix}Z^I& \nabla _\alpha Z^I \end{pmatrix}G_{I\bar J}\begin{pmatrix}\bar Z^{\bar J}\cr \overline\nabla _{\bar \beta }\bar Z^{\bar J}\end{pmatrix}=\begin{pmatrix}-a& 0\cr
 0 & g_{\alpha \bar \beta }\end{pmatrix}
\,.
\label{essential}
\end{equation}
This shows how a basis transformation employing the invertible matrix
\begin{equation}
  \begin{pmatrix}Z^I&\cr \nabla _\alpha Z^I\end{pmatrix}\,,
 \label{invertibleZ}
\end{equation}
relates the superconformal metric $G_{I \bar J}$ to the conical form with projective metric $g_{\alpha \bar \beta}$ (and  radial direction with opposite signature).

A corollary of this relation follows by applying $\nabla _\gamma $ to the second of (\ref{essential}).
The $\nabla _\gamma $ acts only on the factor $\nabla _\alpha Z^I$ from the l.h.s. (using also that $\nabla _\gamma \overline\nabla _{\bar \beta }\bar Z^{\bar J}\propto \bar Z^{\bar J}$ and thus does not contribute using the same relation (\ref{essential})) and gives 0 on the r.h.s.. Since $G_{IJ}$ and the matrix (\ref{ZpaZ}) are invertible, this gives\footnote{Since we now include Levi-Civita connection in the definition of $\nabla _\alpha $ this is the $\widehat{\nabla }$ introduced in footnote~13 in \cite[section17.3.6]{Freedman:2012zz}, and the formula below agrees with that footnote.}
\begin{equation}
  \nabla _\alpha \nabla _\beta Z^I = 0\,,
 \label{nabnab0}
\end{equation}
which will be useful later.

Finally, we can obtain a relation between the curvature tensors of the embedding and projective spaces using an extension of the commutator relations (\ref{commNablagen}) when they act on a quantity with both $\alpha $ and $I$ indices:
\be
[\nabla_\g,\overline\nabla _{\bar \b}] \,\nabla_\a \,Z^I =-a^{-1}g_{\g\bar\b}\nabla _\alpha Z^I -R_{\g\bar \b}{}^\delta {}_\a\nabla_\d\,Z^I +
(y\nabla _\gamma Z^K)(\bar y\overline\nabla _{\bar \beta}\bar Z^{\bar L}) R_{K\bar L}{}^I{}_J\nabla _\alpha Z^J\,.
\label{3derivZ}
\ee
Using (\ref{nabnab0}),  (\ref{commnabZ}) and (\ref{nablaalZ}),  the left hand side is
\begin{equation}
  a^{-1}g_{\alpha \bar \beta  }\nabla _\gamma Z^I\,.
 \label{lhs}
\end{equation}
Next we contract this with $y\bar y G_{I\bar I}\overline\nabla _{\bar \delta }\bar Z^{\bar I} $,  use (\ref{Gg}), and rearrange the furniture to obtain
\begin{equation}
(y\bar y)^2 R_{J\bar IK\bar L}\nabla_\alpha  Z^J\overline\nabla _{\bar \beta } \bar Z^{\bar I} \nabla_\gamma  Z^K\overline\nabla _{\bar \delta }Z^{\bar L}\,=\,R_{\a\bar\beta \gamma \bar\delta }- a^{-1}(g_{\a\bar\beta }g_{\gamma \bar \delta }+  g_{\gamma \bar \beta }g_{\a\bar\delta })\,.
 \label{curvaturerelation}
\end{equation}
Later we will need this relation.

\subsection{Poincar\'{e} fields}

In (\ref{XyZ}),  we related the superconformal coordinates $X^I$ to physical coordinates $z^\alpha$ (and to $y$ which can be considered a relic of the conformal compensator).
In the same spirit, we now introduce the full dictionary between the superconformal components of chiral multiplets ($X^I$, $\O^I$, $\hat F^I$) and super-Poincar\'{e} components ($z^\a$, $\chi^\a$, $h^\a$) (and the superconformal compensator relics $y$, $\chi^0$, $h^0$):
\begin{eqnarray}
 X^I  & = & y  Z^I\,, \nonumber\\
 \Omega^I & = & y \left( Z^I \chi^0 + \nabla _\alpha Z^I \chi^\alpha\right)= X^I \chi^0 + \nabla _\alpha X^I \chi^\alpha\,,\nonumber\\
 \hat F^I &=& y \left( Z^I h^0 + \nabla _\alpha Z^I h^\alpha\right)= X^I h^0 + \nabla _\alpha X^I h^\alpha\,.
 \label{Confvariables}
\end{eqnarray}
The relations of (\ref{essential}) allow us to write the inverse of these relations:
\begin{align}
   \chi^0 & = -a^{-1}\bar y\,\bar Z^{\bar I} G_{\bar IJ}\Omega^J\,,\qquad \chi^\alpha = \bar y\, g^{\alpha \bar \beta }\overline{\nabla }_{\bar \beta }\bar Z^{\bar I}G_{\bar IJ} \Omega^J\,,\nonumber\\
  h^0 & = -a^{-1}\bar y\,\bar Z^{\bar I} G_{\bar IJ}\hat F^J\,,\qquad h^\alpha = \bar y\, g^{\alpha \bar \beta }\overline{\nabla }_{\bar \beta }\bar Z^{\bar I}G_{\bar IJ}\hat F^J\,.
 \label{PoincinConf}
\end{align}
Readers should note how the basis transformation (\ref{invertibleZ}) is used to relate quantities in the superconformal and physical descriptions of the theory.

We choose a gauge fixing for $S$-supersymmetry so that the dilation gauge fixing condition (\ref{dilatationgauge}) is invariant under $Q$-supersymmetry  (\ref{confcovtransf}):
\begin{equation}
S-\mbox{gauge }:\qquad N_I\O^I=G_{I\bar J} \bar X^{\bar J} \O^I=0\qquad \Leftrightarrow  \chi^0 = 0\,.
 \label{Sgaugechi}
\end{equation}
We will address the implications of this gauge choice later on.

\section{Supersymmetry after gauge fixing}
\label{sec:SUSY}
In this section, we discuss the supersymmetry transformations of fields in the projective space  of the super-Poincar\'{e} theory. Due to  gauge fixing, these involve combinations of supersymmetry and other symmetries of the superconformal theory.  These combinations are determined in section \ref{ss:decomp}. They are called Poincar\'e transformations and we use  the notation\footnote{The symbol $\poinc$ comes from point-carr\'{e}.} $\d^\poinc$.  We define covariant transformations on functions of the scalars in section \ref{ss:Poinccovtrans} and denote them by $\hat\d^\poinc$. We discuss their relation with the superconformal supersymmetry, and with the ordinary supersymmetries. On the basis of these properties we define the covariant transformations on other fields in section \ref{ss:covPoincAllfields}. These covariant transformations are far more convenient, and we will only work with them in section \ref{ss:offshellPoincSymm}.
\subsection{Decomposition laws}
\label{ss:decomp}
Since we have gauge-fixed dilatations, $T$-transformations and $S$-supersymmetry, these are no longer independent symmetries of the physical theory.  However, their effects persist in `decomposition laws'  and 'compensating transformations' that express the `gauge-fixed' symmetry parameters in terms of symmetries that remain in the theory, namely $Q$-supersymmetry and K\"{a}hler transformations.

We first consider K\"{a}hler transformations (\ref{Ktrfs}), which leave the dilatation gauge choice (\ref{yybK}) invariant. However, they do not leave the $T$-gauge $y=\bar y$ invariant. This leads to a decomposition law for the $T$-transformations, since the condition transforms under these two symmetries
\begin{equation}
  \left(\delta _{\rm K}[f]+\delta _T[\lambda _T]\right) (y-\bar y) =\frac{1}{a}\left(f(z)y-\bar f(\bar z)\bar y\right)+  \rmi \lambda _T (y+\bar y) \,,
 \label{dely-bary}
\end{equation}
and is only invariant if we include the compensating transformation
\begin{equation}
 \tilde  \lambda _T(f) = \frac{1 }{2a}\rmi (f-\bar f)\,.
 \label{Tinf}
\end{equation}
Therefore, in the Poincar\'{e} theory, K\"{a}hler transformations act on functions $V$ as in (\ref{defcw})
\begin{align}
  \delta^\poinc _{\rm K}[f] V &=
 \left(\delta _{\rm K}[f]+\delta _T[\tilde \lambda _T(f )]\right)V\nonumber\\
 &= \left(-\frac{1}{a}w_+f -\frac{1}{a}w_-\bar f+\rmi c \tilde \lambda _T(f)\right)V = -\frac{1}{a}\left(\hat{w}_+f +\hat{w}_-\bar f\right)V\,.
 \label{PoincKahlertr}
\end{align}
Note that these combine into the same combinations of chiral and \Kahler weights as in (\ref{hatom+-}). In particular, note that though $X^I$ is invariant under the original K\"{a}hler transformations, we have
\begin{equation}
  \delta^\poinc _{\rm K}[f]X^I = \rmi \tilde \lambda _T(f )X^I = -\frac{1 }{2a}(f-\bar f)X^I\,.
 \label{delPoincKX}
\end{equation}

The superconformal transformations are
\begin{equation}
  \delta X^I = \left(\lambda _{\rm D}+\rmi \lambda _T\right)X^I +\frac{1}{\sqrt{2}}\bar \epsilon \Omega ^I\,.
 \label{delXIsconf}
\end{equation}
The combination of these symmetries that remains after gauge fixing should be such that it can be obtained from (\ref{XyZ}) and (\ref{gfT}) in terms of transformations of the Poincar\'{e} variables $\delta z^\alpha $ and $\delta \bar z^{\bar \alpha }$:
\begin{eqnarray}
  \delta X^I &=& (\delta y) Z^I + y\delta Z^I = y\frac{1}{2a}\left(\partial _\alpha {\cal K}\delta z^\alpha +\partial _{\bar \alpha} {\cal K}\delta \bar z^{\bar \alpha}\right)Z^I+ y\partial _\alpha Z^I \delta z^\alpha
  \nonumber\\
  &=& \frac{1}{2a}\left(-\partial _\alpha {\cal K}\delta z^\alpha +\partial _{\bar \alpha} {\cal K}\delta \bar z^{\bar \alpha}\right)yZ^I +y\nabla  _\alpha Z^I\, \delta z^\alpha\,.
 \label{sconfXIz}
\end{eqnarray}
To connect both expressions, we use the decomposition of $\Omega ^I$  in (\ref{Confvariables}), and the gauge (\ref{Sgaugechi}).
Using the invertibility of (\ref{invertibleZ}), we can split the equality of (\ref{delXIsconf}) and (\ref{sconfXIz}) in two equations
\begin{align}
  \lambda _{\rm D}+\rmi \lambda _T = \frac{1}{2a}\left(-\partial _\alpha {\cal K}\delta z^\alpha +\partial _{\bar \alpha} {\cal K}\delta \bar z^{\bar \alpha}\right)  \,, \qquad
   \frac{1}{\sqrt{2}}\bar \epsilon \chi ^\alpha  = \delta z^\alpha \,.
 \label{Poincsusy}
\end{align}
The last one determines that $\chi ^\alpha $ is the supersymmetry partner of $z^\alpha $ in the Poincar\'{e} theory. The first one fixes the dilatation and chiral transformations in terms of the \Poincare supersymmetry variation $\delta z^\alpha$.
We thus obtain that in general the Poincar\'{e} supersymmetry is the following combination of superconformal symmetries
\begin{equation}
  \delta^\poinc [\epsilon ]= \delta _Q[\epsilon ]+\delta _T[\lambda _T(\epsilon )]\,,\qquad
  \lambda _T(\epsilon )= \frac{1}{2a}\rmi\left(\partial _\alpha {\cal K}\delta z^\alpha -\partial _{\bar \alpha} {\cal K}\delta \bar z^{\bar \alpha}\right)\,,
 \label{delPeps}
\end{equation}
involving a compensating chiral transformation (while $\lambda _{\rm D}(\epsilon)=0,$  because the right-hand side of the first equality in (\ref{Poincsusy}) is imaginary).

\subsection{Poincar\'{e} covariant transformations on functions of scalars}
\label{ss:Poinccovtrans}
While compatible with the \Poincare gauge choices, the above do not yet constitute covariant supersymmetry transformations.
We \emph{define} the covariant derivatives of functions of $z$ and $\bar z$ as a generalization of (\ref{nablaconfscalar}):
\begin{equation}
  \hat{\delta }^\poinc[\epsilon ]V(z,\bar z)= (\nabla _\alpha V)\delta z^\alpha + (\overline{\nabla }_{\bar \alpha }V)\delta \bar z^{\bar \alpha }\,,
 \label{hatdeltatonab}
\end{equation}
using the covariant derivatives defined in section \ref{ss:covderproj}. We apply this also if $V$ contains other indices $I, \, \alpha ,\ldots $; in this case the appropriate connections $\G^I_{JK},~\G^{\alpha}_{\beta\gamma}$ are included in the $\nabla _\alpha $ and $\overline{\nabla }_{\bar \alpha }$ operation as in (\ref{defnablaVI}) and  (\ref{nablaaVbeta}).

Observe immediately the difference with ordinary transformations which follow the normal chain rule:
\begin{equation}
  \delta^\poinc [\epsilon ]V(z,\bar z)= (\partial _\alpha V)\delta z^\alpha + (\overline{\partial }_{\bar \alpha }V)\delta \bar z^{\bar \alpha }\,.
 \label{delchainrule}
\end{equation}

Consider first the action of (\ref{hatdeltatonab}) on the conformal scalars $X^I$:
\begin{equation}
  \hat{\delta}^\poinc[\epsilon ] X^I=\nabla  _\alpha X^I\, \delta z^\alpha= \frac{1}{\sqrt{2}}\nabla  _\alpha X^I\,\bar \epsilon \chi ^\alpha= \frac{1}{\sqrt{2}}\bar \epsilon\Omega ^I=
\delta _Q[\epsilon ]X^I=  \hat\delta _Q[\epsilon ]X^I \,,
 \label{delPoincXcov}
\end{equation}
where for the third equality we used (\ref{Confvariables}) with $\chi ^0=0$ due to the $S$-gauge condition. We thus obtain that this covariant superconformal transformation on this field is equal to
the $Q$-transformation of the superconformal theory.
This is also true for all other covariant objects that are well-defined in the superconformal theory. For these, we can use (\ref{covchainnabla}), and thus
\begin{equation}
 \hat{\delta}^\poinc[\epsilon ] V^I=\delta z^\alpha\nabla  _\alpha X^J\nabla _J V^I+ \delta \bar z^{\bar \alpha}\overline{\nabla}  _{\bar \alpha} \bar X^{\bar J}\overline{\nabla} _{\bar J} V^I=
 \frac{1}{\sqrt{2}}\bar \epsilon \left[\Omega ^J \nabla _J V^I+ \Omega ^{\bar J}\overline{\nabla} _{\bar J} V^I\right]=\hat{\delta }_Q V^I\,,
 \label{hatdeltaequal}
\end{equation}
according to (\ref{nablaconfscalar}) for the conformal $Q$-supersymmetry.

Notice that these equalities were not valid for the non-covariant transformations, see (\ref{sconfXIz}), which lead to a compensating $T$-transformation in (\ref{delPeps}).
The covariant transformations leave the gauge conditions for dilatations and  $T$-transformation invariant. The covariant transformation of the dilatation condition $N=-a$ is according to (\ref{hatdeltaequal}) the same as the $Q$ transformation:
\begin{equation}
 \hat\delta ^\poinc[\epsilon ]N = \frac{1}{\sqrt{2}}\bar{\epsilon} \left(N_I\Omega ^I + N_{\bar I}\Omega ^{\bar I}\right)=0\,,
 \label{covtransdilat}
\end{equation}
due to the $S$-gauge condition (\ref{Sgaugechi}). Thus the compensating dilatation transformation vanishes, i.e. $\lambda _{\rm D}(\epsilon)=0$.

The next gauge condition is the $T$-gauge $y=\bar y$. But since $y$ is covariantly constant, see (\ref{nablaay0}),
\begin{equation}
  \hat{\delta }^\poinc[\epsilon ] y = (\nabla_\a y)\d z^\a +(\overline\nabla _{\bar\a}y)\d \bar z^{\bar\a} =0\,.
 \label{hatdely0}
\end{equation}
Therefore, the $T$-gauge condition $y=\bar y$ is invariant and the decomposition law for the covariant transformations is $\lambda _T(\epsilon )=0$, to be confronted with (\ref{delPeps}) for the usual super-Poincar\'{e} transformations.

For completeness, we now discuss the relation between these transformations and the ordinary transformations. Comparing (\ref{hatdeltatonab}) and  (\ref{delchainrule}), we find with (\ref{defnablaV}):
\begin{equation}
  \hat{\delta }^\poinc[\epsilon ]V = \delta^\poinc[\epsilon ] V  +a^{-1}\left[\hat{w}_+(\partial _\alpha {\cal K})\delta z^\alpha + \hat{w }_-(\partial _{\bar \alpha} {\cal K})\delta \bar z^{\bar \alpha }\right] V\,.
 \label{hatdeldelV}
\end{equation}
For objects already defined in the superconformal theory, $\hat{w}_+=-\hat{w}_-= \ft12c$, and the correction term just amounts to the $T$-compensating transformation in (\ref{delPeps}).
The Poincar\'{e} theory involves also quantities that are defined only after the split of variables  (\ref{XyZ}). For example, this applies to the superpotential $W$ that is defined from the superconformal ${\cal W}$ by ${\cal W}=y^3 W$. For these, we do not have the conformal transformations. Their covariant Poincar\'{e} transformations are related to the ordinary Poincar\'{e} transformations by the full rule (\ref{hatdeldelV}).

As an example we apply (\ref{hatdeldelV}) to the superpotential $W$, which has $\hat{w}_+=3$ and $\hat{w}_-=0$:
\begin{equation}
\begin{split}
  \hat{\delta }^\poinc[\epsilon ]W &= \delta^\poinc[\epsilon ] W +3a^{-1}(\partial _\alpha {\cal K}) W\delta z^\alpha\\
  &= \partial _\alpha W\,\delta z^\alpha +3a^{-1}(\partial _\alpha {\cal K})\, W\delta z^\alpha = \nabla _\alpha W\, \delta z^\alpha \,.
\end{split}
 \label{applytoW}
\end{equation}

For quantities with various indices the relation between the covariant and ordinary Poincar\'{e} covariant derivatives is as in
\begin{eqnarray}
 \hat{\delta }^\poinc[\epsilon ]V^I & \equiv  & \delta^\poinc[\epsilon ] V^I+\frac1a V^I\left[\hat{w}_+(\partial _\alpha {\cal K})\delta z^\alpha + \hat{w }_-(\partial _{\bar \alpha} {\cal K})\delta \bar z^{\bar \alpha }\right]+\Gamma ^I_{JK}V^J\nabla _\beta  X^K\delta z^\beta  \,,\nonumber\\
\hat{\delta }^\poinc[\epsilon ]V^\gamma &\equiv & \delta^\poinc[\epsilon ] V^\gamma+\frac1a V^\gamma \left[\hat{w}_+(\partial _\alpha {\cal K})\delta z^\alpha + \hat{w }_-(\partial _{\bar \alpha} {\cal K})\delta \bar z^{\bar \alpha }\right] + \Gamma _{\alpha \beta }^\gamma V^\alpha \delta z^\beta  \,.
 \label{hatdelVPoinc}
\end{eqnarray}

\subsection{Poincar\'{e} covariant transformations on all fields}
\label{ss:covPoincAllfields}

Now we consider also the other fields in the theory, which are the fermions and auxiliary fields from the chiral multiplets, and the fields of the Weyl multiplet.
We saw above that for the scalars the covariant Poincar\'{e} supersymmetry can be identified with the covariant $Q$-supersymmetry of the conformal theory. However, we still have to consider the $S$-gauge condition, which is not invariant under the $Q$-supersymmetry. Therefore we define the covariant Poincar\'{e} supersymmetry as the combination of covariant $Q$ and $S$ transformations that preserve the gauge fixing\footnote{If there are objects that transform under special conformal transformations, we furthermore have to use the decomposition law following from the gauge condition $b_\mu =0$ :
 $ \lambda _{{\rm K}\,\mu} = -\ft14 \bar \epsilon \phi _\mu +\ft14 \bar \psi _\mu \eta (\epsilon )$.}
\begin{equation}
  \hat{\delta}^\poinc[\epsilon ] =\hat{\delta} _Q[\epsilon ]+ \hat{\delta}_S [\eta (\epsilon )] \,.
 \label{hatPoincisQS}
\end{equation}
The decomposition law the $S$ supersymmetry, $\eta (\epsilon )$, will be obtained in section \ref{ss:transfolaws}, see (\ref{Sdecomposition}).

The covariant superconformal transformations leave the action invariant. The super-Poincar\'{e} action is defined as the superconformal action after applying gauge conditions. Since the covariant super-Poincar\'{e} transformations leave these gauge conditions invariant and are a linear combination of the superconformal ones, they must also leave the action invariant.

\section{Off-shell Poincar\'e supersymmetry}
\label{ss:offshellPoincSymm}

In this section  we will derive and present our main results: the full supersymmetry action and transformation laws of the super-\Poincare theory including the auxiliary fields of the chiral multiplets. But let us first summarize where we stand.

The previous section involved some subtle arguments, but the upshot is very simple.
The covariant supersymmetry transformations of the superconformal theory reduce to the covariant supersymmetry transformations of the Poincar\'{e} theory after gauge fixing is properly incorporated. Therefore, we will remove the indication $^\poinc$ on $\hat{\delta }$ from now on.

These covariant Poincar\'{e} symmetry transformations act in a simple way on functions of $z$ and $\bar z$: we have the covariant expressions (\ref{nablaVsplit}-\ref{defnablaV}) and (\ref{hatdeltatonab}):
\begin{equation}
  \nabla _\mu V(z,\bar z)= \left(\partial _\mu z^\alpha \nabla _\alpha + \partial _\mu \bar z^{\bar \alpha} \nabla _{\bar \alpha} \right)V \,,\qquad
  \hat{\delta }V(z,\bar z)= \left(\delta  z^\alpha \nabla _\alpha + \delta  \bar z^{\bar \alpha} \nabla _{\bar \alpha} \right)V \,.
 \label{rulesVzbarz}
\end{equation}
These relations also hold if $V$ carries indices $I$ or $\alpha $.
Further the covariant derivatives (and thus covariant transformations) vanish on $y$ and on the metrics, e.g.
\begin{equation}
  \hat{\delta }y=0\,,\qquad \hat{\delta }G_{I\bar J}=0\,,\qquad \hat{\delta }g_{\alpha \bar \beta }=0\,.
 \label{vanishingcovder}
\end{equation}
In fact, $y$ is totally inert: $\nabla_\m y = \nabla_\a y=\nabla_{\bar\a}y =0$.
Thus one can use the definition $X^I=yZ^I(z)$, and
\begin{equation}
  \nabla _{\bar \alpha }Z^I= 0\,,\qquad \nabla _\alpha \nabla _\beta Z^I=0\,,\qquad \nabla _{\bar \alpha }\nabla _\beta Z^I=a^{-1}g_{\beta\bar \alpha}Z^I\,,
 \label{simplederZ}
\end{equation}
and in general commutators of covariant derivatives give curvatures in the embedding and projective spaces, related by (\ref{curvaturerelation}). We will frequently use the metric relations encoded in the matrix equation (\ref{essential}).

Finally, we will often need the covariant transformation of $\nabla _\alpha X^I$,
\begin{equation}
  \hat{\delta }\nabla _\alpha X^I= \nabla _\beta \nabla _\alpha X^I \delta z^\beta  +\nabla _{\bar \beta} \nabla _\alpha X^I \delta z^{\bar \beta}= a^{-1} X^I g_{\alpha \bar \beta }\delta \bar z^{\bar \beta }\,,
 \label{hatdelnabX}
\end{equation}
which follows from (\ref{rulesVzbarz}) and (\ref{simplederZ}).

\subsection{Transformation laws}
\label{ss:transfolaws}

An integral part of  the covariant rules in section \ref{ss:susytransfcov} are the covariant derivatives of scalars and fermions \eqref{covderconf} and (\ref{wcD}). These can be rewritten as
\begin{eqnarray}
 {\cal D}_\mu X^I & =& \nabla _\alpha X^I \, \partial _\mu z^\alpha -\rmi A_\mu ^{\rm F} X^I -\frac{1}{\sqrt{2}}\bar \psi _\mu \Omega ^I\,,\nonumber\\
 \widehat{{\cal D}}_\mu \Omega ^I&=&\nabla _\mu \Omega ^I+\frac{1}{2}\rmi A_\mu ^{\rm F}\Omega ^I-\frac{1}{\sqrt{2}}P_L\left(\slashed{\cal D}X^I+\hat F^I\right)\psi _\mu -\sqrt{2}P_LX^I\phi _\mu \,.
 \label{covdervfromblackboard}
\end{eqnarray}
where we have split the auxiliary gauge field of the $T$-symmetry into its bosonic and fermionic part:
\begin{equation}
  A_\mu ={\cal A}_\mu + A_\mu ^{\rm F}\,,\qquad
  A_\mu ^{\rm F}=\rmi\frac{1}{4N}\left[\sqrt{2}\bar \psi_\mu\left(N_I
\Omega^I-N_{\bar I} \Omega ^{\bar I} \right)+N_{I\bar J}\bar \Omega ^I\gamma _\mu \Omega
^{\bar J}\right]  \,,
 \label{Amusplit}
\end{equation}
and ${\cal A}_\mu $ has the value in (\ref{calAmualpha}). The latter has been used to define
\begin{equation}
  \nabla _\mu \Omega ^I = \left(\partial _\mu +\ft14 \omega _\mu {}^{ab}\gamma _{ab}+ \ft12 \rmi{\cal A}_\mu \right)\Omega ^I+ \Gamma ^I_{JK}\Omega ^J \nabla _\mu X^K\,.
 \label{nablaOmega}
\end{equation}

In order to express these in terms of \Poincare variables, we use the relations \eqref{Confvariables} to expand all fields in terms of the basis $(X^I, \nabla_\alpha X^I)$. Moreover, we will use the gauge condition $\chi^0 = 0$. To illustrate this procedure, let us focus on the expression for $A_\mu ^{\rm F}$. The first term vanishes by the $S$-gauge condition (\ref{Sgaugechi}). In the second term, we express fermions in terms of the basis and find
 \begin{align}
 A_\mu ^{\rm F}= - \frac{1}{4a}\rmi N_{I\bar J}\bar \Omega ^I\gamma _\mu \Omega^{\bar J} = - \frac{1}{4a}\rmi N_{I \bar J} \nabla_\alpha X^I \nabla_{\bar \beta} X^{\bar J} \bar \chi^\alpha \gamma_\mu \chi^{\bar \beta} = - \frac{1}{4a}\rmi g_{\alpha \bar \beta} \bar \chi^\alpha \gamma_\mu \chi^{\bar \beta} \,,
 \label{AmuFpoinc}
 \end{align}
where we have used \eqref{essential}.

The full covariant derivative of the scalars in the $(X^I, \nabla_\alpha X^I)$ basis reads
 \begin{align}
  {\cal D}_\mu X^I = \nabla_\alpha X^I {\cal D}_\mu z^\alpha -\rmi A_\mu ^{\rm F} X^I \,, \qquad {\cal D}_\mu z^\alpha = \partial_\mu z^\alpha - \frac{1}{\sqrt{2}} \bar \psi_{\mu} \chi^\alpha \,,
  \label{covdersc}
 \end{align}
where we have defined the supercovariant derivative on \Poincare scalars. Similarly, the covariant derivative of the fermions is given by
 \begin{align}
 \hat {\cal D}_\mu \Omega^I & = \nabla_\alpha X^I \left(\hat {\cal D}_\mu \chi^\alpha + \ft{1}{2}\rmi A_\mu^{\rm F} \chi^\alpha\right) + X^I \left( a^{-1} g_{\alpha \bar \beta} \chi^\alpha {\cal D}_\mu z^{\bar \beta} - \sqrt{2} P_L \hat \phi_\mu \right) \,,
\end{align}
in terms of the \Poincare covariant derivatives
 \begin{align}
  \hat {\cal D}_\mu \chi^\alpha & = \nabla_\mu \chi^\alpha - \frac{1}{\sqrt{2}} P_L (\slashed {\cal D} z^\alpha +h^\alpha) \psi_\mu \,, \qquad
  P_L\hat \phi_\mu  = P_L\phi_\mu + \frac12 P_L\left(h^0 -3\rmi \slashed A^{\rm F}\right) \psi_\mu\,,\nonumber\\
  \nabla_\mu \chi^\alpha &= \left(\partial _\mu +\ft14\omega _\mu {}^{ab}\gamma _{ab}+ \ft32 \rmi{\cal A}_\mu \right)\chi ^\alpha +\Gamma ^\alpha _{\beta \gamma }\chi ^\beta \partial _\mu z^\gamma \,.
  \label{covariantphi}
 \end{align}
We will see below why the expression for $\hat \phi_\mu$ is a good Poincar\'{e} covariantization of $\phi _\mu $.

We now turn to the supersymmetry transformations, starting from their conformal counterparts \eqref{confcovtransf}. We first repeat the derivation of the transformations of the scalars in (\ref{Poincsusy}) in a more direct way, illustrating the covariant methods:
\be
\hat\delta X^I = \frac{1}{\sqrt2} \bar\e P_L\O^I= \frac{1}{\sqrt2} \bar\e P_L\nabla_\a X^I\chi^\a =  \nabla_\a X^I \d z^\a \,,
\ee
using \eqref{hatdeltatonab}, from which it readily follows that
\begin{equation}
  \delta z^\alpha = \frac{1}{\sqrt{2}}\bar \epsilon\chi ^\alpha \,.
 \label{delz}
\end{equation}

Covariant methods allow us to find the transformations of the other Poincar\'{e} fields quite easily. Let us start with the fermion transformation. Expanding both the superconformal transformations and the transformations of the \Poincare fields in terms of the covariant basis, one has
 \begin{align}
  \hat\delta \Omega^I & = X^I \hat\delta \chi^0 + \nabla_\alpha X^I \hat\delta \chi^\alpha + \hat\delta(\nabla_\alpha X^I) \chi^\alpha  \,, \notag \\
 & = X^I (\hat\delta \chi^0 + \tfrac1a g_{\alpha \bar \beta} \chi^\alpha \delta \bar z^{\bar \beta} ) + \nabla_\alpha X^I \hat\delta \chi^\alpha \,.
 \end{align}
On the other hand, from the conformal transformation in (\ref{confcovtransf}) using the covariant derivative \eqref{covdersc} discussed above, we get
\begin{equation}
  \hat\delta \Omega^I=\frac{1}{\sqrt{2}} P_L \left[ \nabla_\alpha X^I ( \slashed {\cal D} z^\alpha + h^\alpha)\epsilon  + X^I \left( -\rmi \slashed{A}^{\rm F}
  + h^0  \right)\right]\epsilon  +\sqrt{2}P_L\eta X^I \,.
 \label{hatdelOmfromconf}
\end{equation}
Equating these two expressions for the superconformal transformations leads to the Poincar\'{e} transformations for the fermions,
 \begin{align}
 \hat\delta \chi^\alpha = \frac{1}{\sqrt{2}} P_L (   \slashed {\cal D} z^\alpha + h^\alpha) \epsilon \,,\qquad
   \sqrt{2} \hat\delta \chi^0 =P_L\left(h^0 - 3 \rmi \gamma^\mu A_\mu ^{\rm F}\right)\epsilon
 +2 P_L\eta\,,
 \label{hatdelchi}
\end{align}
where we have performed a Fierz rearrangement to bring both fermion trilinear expressions to the same form.

In addition to the transformation of the physical fermion, (\ref{hatdelchi}) also leads to the decomposition law announced in (\ref{hatPoincisQS}).
The requirement that the SUSY variation of the remaining component $\chi^0$ leaves the gauge choice $\chi^0 = 0$ invariant, i.e.~$\hat \delta \chi^0 = 0$, results in the compensating transformation of the gauge fixed $S$-supersymmetries:
 \begin{align}
   P_L \eta(\epsilon ) 
   =\ft12   P_L\left(-  h^0  +3\rmi \slashed{A}^{\rm F}\right)\epsilon \,.
   \label{Sdecomposition}
  \end{align}

Finally, the SUSY transformation laws of the auxiliary fields can be derived in the same way. On the one hand,  from a covariant transformation on the definition in (\ref{Confvariables}), we have the expression
 \begin{align}
  \hat\delta \hat F^I = X^I  \left(\hat\delta h^0 + \frac1a g_{\alpha \bar \beta} \delta z^{\bar \beta} h^\alpha\right) + \nabla_\alpha X^I  \left( \hat\delta h^\alpha + \frac{1}{\sqrt{2}} h^0 \bar \epsilon \chi^\alpha \right) \,,
 \end{align}
while the superconformal transformation laws are expanded into (using (\ref{curvaturerelation}))
 \begin{align}
 \sqrt{2}\hat\delta \hat F^I = & \nabla_\alpha X^I \bar \epsilon \left[ \hat{\slashed{\cal D}}\chi^\alpha + \ft{1}{2}\rmi \slashed{A}^F \chi^\alpha) + \ft12\left(R_{\beta \bar \gamma}{}^\alpha {}_{\delta} + 2a^{-1} g_{\beta \bar \gamma} \delta_\delta^\alpha \right)  \chi^{\bar \gamma} \bar \chi^\beta \chi^{\delta} \right] + \notag \\
& +  X^I \bar \epsilon\left[  a^{-1} g_{\alpha \bar \beta} {\slashed {\cal D}} z^{\bar \beta } \chi^\alpha   - \sqrt{2} P_R \gamma ^\mu \hat \phi_\mu \right] \,.
 \end{align}
This leads to the transformation laws for the \Poincare auxiliary fields
 \begin{align}
\sqrt{2} \hat \delta h^0 & =  \bar \epsilon \left[ a^{-1} g_{\bar \alpha \beta}{\slashed {\cal D}} z^{\bar \alpha} \chi^\beta  - \sqrt{2} P_R\gamma ^\mu  \hat \phi_\mu)  - a^{-1} g_{\alpha \bar \beta} \chi ^{\bar \beta} h^\alpha \right]\,, \nonumber\\
 \sqrt{2}\hat \delta h^\alpha & =  \bar \epsilon ( \hat{\slashed {\cal D}} \chi^\alpha + \ft12 \rmi\slashed A^F \chi^\alpha ) -  h^0 \bar \epsilon \chi^\alpha + \ft12\left(R_{\beta \bar \gamma}{}^\alpha {}_{\delta} +2a^{-1}g_{\beta \bar \gamma} \delta_\delta^\alpha \right) \bar \epsilon \chi^{\bar \gamma} \bar \chi^\beta \chi^{\delta} \nonumber\\
 &= \bar \epsilon \left[\left( \hat{\slashed {\cal D}}  + \ft32 \rmi\slashed A^F  -  h^0\right)  \chi^\alpha + \ft12R_{\beta \bar \gamma}{}^\alpha {}_{\delta}  \chi^{\bar \gamma} \bar \chi^\beta \chi^{\delta}\right]\,.\label{hatdeltah}
 \end{align}

The covariant SUSY transformations above are amongst the central results of this paper. The covariant transformations of the fermions have an additional contribution due to the off-shell auxiliary fields $h^\alpha$. Finally, we also obtained the transformation of the auxiliary fields \eqref{hatdeltah}. Note that these contain specific quartic fermion terms, which organize into \Kahler curvatures by virtue of covariance.

We now consider fields of the Weyl multiplet. At the superconformal level these fields and their  covariant superconformal transformations are independent of the K\"{a}hler target space.
However, we have to take the decomposition laws into account, which differ for ordinary and covariant Poincar\'{e} transformations. The transformation of the frame field is not affected and reads:
\begin{equation}
  \delta e_\mu ^a = \ft12\bar \epsilon \gamma ^a\psi _\mu \,.
 \label{deltaemua}
\end{equation}
The field $b_\mu $ is eliminated by the dilatation gauge choice, and $A_\mu $ has been replaced by ${\cal A}_\mu + A^F_\mu$. The covariant transformation of the gravitino remains; for this we need the decomposition law (\ref{hatPoincisQS}). Furthermore, we will apply the split (\ref{Amusplit}), and include the bosonic part in a covariant derivative:
\begin{equation}
\begin{split}
  \hat{\delta }P_L\psi _\mu &= \hat\delta _Q\psi _\mu -\gamma _\mu \eta (\epsilon )\\
   &= \nabla _\mu P_L\epsilon -\ft32\rmi A_\mu ^{\rm F}P_L\epsilon +\ft12\gamma _\mu P_R\left( \bar h^0  +3\rmi \slashed{A}^{\rm F}\right)\epsilon \,,\\
  \nabla _\mu P_L\epsilon &= \left(\partial  _\mu +\ft14\omega _\mu {}^{ab}\gamma _{ab}-\ft32\rmi {\cal A}_\mu \right)P_L\epsilon\,.
  \end{split}
 \label{covtransfgravitino}
\end{equation}
Note that $\epsilon $ can be considered as having chiral weight $c=3/2$.

Due to the compensating $S$-transformation, the super-Poincar\'{e} covariant gravitino curvature also receives a contribution (compare with\cite[(11.72)]{Freedman:2012zz}
\begin{equation}
  P_R\hat R_{\mu \nu }^\poinc= P_R R'_{\mu \nu }(Q)+2\gamma _{[\nu }\eta (\psi_{\mu ]})= P_R R'_{\mu \nu }(Q)- P_R \gamma _{[\nu }\left(h^0-3\rmi \slashed{A}^{\rm F}\right) \psi _{\mu ]}\,.
 \label{hatRprime}
\end{equation}
Note that $\hat{\phi }_\mu $ in (\ref{covariantphi}) is related to $\hat R_{\mu \nu }^\poinc$ in the same way as $\phi _\mu $ is defined from $R'_{\mu \nu }(Q)$ in (\ref{valuephimu}).

\subsection{Superpotential part of the action}
\label{ss:PoincAct}

To write the action in Poincar\'{e} variables, is now a straightforward substitution using the new variables. Instead of deriving the full action in this paper, we will only highlight those parts that contain the auxiliary fields. The remainder of the action coincides with the on-shell version and can be found in \cite{Freedman:2012zz}.

As an example of the calculation of the action, we consider the $F$-term of the geometric formulation of the superconformal theory (\ref{WF}):
\begin{equation}
  [\mathcal{W}]_Fe^{-1}=
{\cal W}_I \hat F^I -\ft12{\cal  W}_{I;J}\bar \Omega^{I}\Omega ^{J}
 +\frac{1}{\sqrt{2}}{\cal  W}_I\bar \psi\cdot
\gamma  \Omega^{I}+\ft12 {\cal  W}\bar \psi _{\mu }P_R \gamma ^{\mu \nu
}\psi _{\nu }+ \hc\,,
 \label{WFrepeat}
\end{equation}
We will use the relation between conformal and Poincar\'{e} fields as in (\ref{Confvariables}), but we set $\chi^0=0$ due to the $S$-gauge condition. The superpotential ${\cal W}$ obeys two fundamental relations: one is the definition of the Poincar\'{e} superpotential $W$, and the other is the statement that it should be homogeneous of degree 3:
\begin{equation}
  {\cal W}=y^3 \, W\,,\qquad  y\,Z^I {\cal W}_I= 3{\cal W}=3y^3 W\,.
 \label{fundamW}
\end{equation}
Taking covariant derivatives leads to two new equations
\begin{equation}
  {\cal W}_I\, \,\nabla _\alpha Z^I = y^2\, \nabla _\alpha W\,,\qquad \nabla _\alpha Z^I\,{\cal W}_I +y\,Z^I{\cal W}_{I;J}\nabla _\alpha Z^J= 3y^2\nabla _\alpha W\,.
 \label{1stderfundW}
\end{equation}
We further need the second derivative of the first equation, using (\ref{nabnab0}):
\begin{equation}
  {\cal W}_{I;J}\, \,\nabla _\alpha Z^I\,\nabla _\beta Z^J = y\nabla _\alpha\nabla _\beta  W\,.
 \label{2ndderfundW}
\end{equation}
This easily leads to
\begin{eqnarray}
  [\mathcal{W}]_Fe^{-1}&=&
y{\cal W}_I \left( Z^I h^0 + \nabla _\alpha Z^I h^\alpha\right) -\ft12y^2{\cal  W}_{I;J}\nabla _\alpha Z^I \bar \chi^\alpha \nabla _\beta  Z^J\chi ^\beta  \nonumber\\
&& +\frac{1}{\sqrt{2}}y{\cal  W}_I\bar \psi\cdot
\gamma   \nabla _\alpha Z^I \chi^\alpha+\ft12 {\cal  W}\bar \psi _{\mu }P_R \gamma ^{\mu \nu
}\psi _{\nu }+ \hc\nonumber\\
&=&y^3\left[3\, W\,h^0 + \nabla _\alpha W\, h^\alpha -\ft12\nabla _\alpha \nabla _\beta W \bar \chi^\alpha \chi ^\beta\right.  \nonumber\\
&& \left.+\frac{1}{\sqrt{2}}\nabla _\alpha W\bar \psi\cdot
\gamma   \chi^\alpha+\ft12 W\bar \psi _{\mu }P_R \gamma ^{\mu \nu
}\psi _{\nu }\right]+ \hc\,,
 \label{WFPoinc}
\end{eqnarray}
where $y^3 = \rme^{\kappa ^2{\cal K}/2}$.

\subsection{The auxiliary field action}
\label{ss:auxfieldAction}

The $[{\cal W}]_F$ action of (\ref{WF}) contains a linear term in the ${\hat F}^I$, while $[{\cal N}]_D$ includes a quadratic term.  They can be combined in the quadratic expression:
\begin{equation}
e^{-1} {\cal L}_F=G_{I\bar J} ( \hat F^I -\hat{F}^I_{G} ) ( \bar {\hat{F}}^{\bar J}- \bar{\hat{F}}^{\bar J}_G)-G_{I\bar J}\hat{F}^I_{G}\bar{\hat{F}}^{\bar J}_G\,,\qquad
\hat{F}_G^I= -G^{I\bar J} \overline{{\cal W}}_{\bar J}\,.
\label{LFinIJ}
\end{equation}
To express this in Poincar\'{e} variables, the formula in \cite[(17.85)]{Freedman:2012zz} is convenient:
\begin{equation}
  G^{I\bar J}=y\bar{y}\left( -\ft1aZ^I\bar Z^{\bar J}
  +g^{\alpha \bar \beta }\nabla _\alpha Z^I\,\overline\nabla _{\bar \beta }\bar Z^{\bar J}\right) \,.
\label{inverseG}
\end{equation}
Using the definitions in  (\ref{Confvariables}) as well as the gauge-fixing conditions, the auxiliary action for $h^0$ and $h^\alpha$ becomes
\begin{equation}
e^{-1} {\cal L}_F=g_{\alpha \bar \alpha} (h^\alpha - h^\alpha_G) (\bar h^{\bar \alpha} - \bar h^{\bar \alpha}_G) -  g_{\alpha \bar \alpha}  h^\alpha_G \bar h^{\bar \alpha}_G - a (h^0 - h^0_G) (\bar h^{\bar 0} - \bar h^{\bar 0}_G) + 3  h^0_G  \bar h^{\bar 0}_G \,.
\end{equation}
Here  the Gaussian values are given by
\begin{equation}
h^\alpha_G= - g^{\alpha \bar \beta } \bar y^3\overline{\nabla}_{\bar \beta }\overline{W} \,,\qquad h^0_G=\frac{3}{a}\bar y^3\overline{W}\,.\label{feh}
\end{equation}
 We will keep both $h^{\alpha}$ and $h^0$ off shell for the future work with a nilpotent multiplet.\footnote{We are not retaining other auxiliary fields like $A_\mu$ and $D^A$ since they are not affected by the constraints on chiral multiplets and take their Gaussian values, as shown in \cite[(17.21)]{Freedman:2012zz}.} The result for the full auxiliary field Lagrangian is then
\begin{equation}\label{action}
e^{-1} {\cal L}=e^{-1} {\cal L}_{\rm book} + g_{\alpha \bar \alpha} (h^\alpha - h^\alpha_G) (\bar h^{\bar \alpha}- \bar h^{\bar \alpha}_G) - a (h^0 - h^0_G) (\bar h^{\bar 0} - \bar h^{\bar 0}_G) \,,
\end{equation}
where ${\cal L}_{\rm book}$ is the result in\cite[section 18.1]{Freedman:2012zz}.  The quadratic term in $h^\alpha $ also appears in \cite{Kallosh:2015tea}.

In a theory in which there are algebraic constraints on chiral multiplets,  one can eliminate the auxiliary fields
$h^\alpha$ by means of an order by order  expansion in fermion bilinears about their Gaussian values:
\begin{equation}
h^\alpha_{\rm on-shell}= h^\alpha_G + \Delta h^\alpha\,.
\end{equation}
The expansion terminates because of  Grassmann anti-commutation.

\subsection{Consistency}

Having derived the full action with auxiliary fields, as well as their transformation laws, a consistency check is to prove that these transformation laws leave the on-shell values \eqref{feh} invariant; in other words, is integrating out the auxiliary fields compatible with unbroken supersymmetry or not? Of course this question will be answered affirmatively (otherwise there would not be supergravity theories with on-shell auxiliary fields), but in reaching this conclusion we will gain an understanding of the structure of the supersymmetry transformations of the auxiliary fields. In particular, we will show that they take a very simple on-shell form.

Which field equations should we use? The transformations of the auxiliary fields $h^\alpha$ of the matter multiplets are expected to be related to the field equations of their fermionic partners $\chi ^\alpha $. Indeed, from e.g. \cite[(18.6)]{Freedman:2012zz} we can derive the field equation of these fermions, and after recombining various explicit gravitino terms in covariant derivatives, one finds indeed an expression related to (\ref{hatdeltah}). This leads to
 \begin{align}
 \sqrt{2}\hat \delta h^{\alpha} & =-g^{\alpha \bar \beta }e^{-1}\frac{\delta {\cal L}}{\delta \bar \chi ^{\bar \beta }} - h^0 \bar \epsilon  \chi^{\alpha} - g^{\alpha \bar \beta }\overline{m}_{\bar \beta \bar \gamma }\chi ^{\bar \gamma }+\frac{1}{\sqrt{2}}\gamma \cdot \psi (h_G^\alpha -h^\alpha )\nonumber\\
 &\approx  -3a^{-1}y^3\overline{W}\bar \epsilon \chi ^\alpha -y^3 g^{\alpha \bar \beta } \overline{\nabla }_{\bar \gamma }\overline{\nabla }_{\bar \beta  }\overline{W}\bar \epsilon \chi ^{\bar \gamma }= \sqrt{2}\hat \delta h^{\alpha}_G\,,
\end{align}
where the use of field equations is indicated by $\approx $.
This thus coincides with the covariant transformation of the on-shell value of $h^\alpha$.

The partner of the auxiliary field $h^0$ was $\chi ^0$, but this field has already been eliminated by the $S$-supersymmetry gauge condition. However, we can trace back that field equation to another one by recalling that the conformal action $S_{\rm conf}$ was invariant under $S$-supersymmetry. The only fields that transform under $S$-supersymmetry are the gravitino and $\chi ^0$, see (\ref{hatdelchi}), and thus the invariance of the action under $S$-supersymmetry is the statement
\begin{equation}
  0=\delta_S S_{\rm conf} =\frac{\delta  S_{\rm conf}}{\delta \chi ^0} \delta _S\chi ^0 +  \frac{\delta  S_{\rm conf}}{\delta \psi _\mu }(-\gamma _\mu \eta) \,.
 \label{Sinvariancechi0}
\end{equation}
Hence the field equation of $\chi ^0$ is proportional to the trace of the field equation of the gravitino. Therefore we start with the gravitino field equation obtained from \cite[(18.6)]{Freedman:2012zz}:
\begin{eqnarray}
 P_L\Sigma ^\mu  & \equiv  & e^{-1}P_L\frac{\delta S}{\delta \bar \psi _\mu } \nonumber\\
   & = &
  -\kappa ^{-2}P_L\gamma ^{\mu \nu \rho }\left(\partial _\nu +\ft14\omega _\nu {}^{ab}\gamma _{ab}+\ft32\rmi A_\nu \right)\psi _\rho +\frac{1}{\sqrt{2}}g_{\alpha \bar \beta }{\slashed{\cal D }}\bar z^{\bar \beta }\gamma ^\mu \chi ^\alpha \nonumber\\
   &&+\kappa ^2y^3 P_L\gamma ^{\mu \nu }\psi _\nu \overline{W}+\frac{1}{\sqrt{2}}\gamma^\mu y^3\chi ^{\bar \alpha} \nabla _{\bar \alpha}\overline{W}+3\rmi P_L\gamma ^{\nu \mu }\slashed{A}^{\rm F}\psi _\nu    \,.  \label{gravfe}
\end{eqnarray}
Its contraction   (using \cite[(16.26)]{Freedman:2012zz}) is
\begin{align}
   P_R\gamma ^\mu \Sigma _\mu=&6\kappa ^{-2}P_R\gamma ^\mu \hat\phi _\mu -\sqrt{2}g_{\alpha \bar \beta }{\slashed{\cal D}}\bar z^{\bar \beta }\chi ^\alpha+ 3P_R\gamma \cdot \psi (h^0_G-h^0)+2\sqrt{2}y^3\chi ^{\bar \alpha} \nabla _{\bar \alpha}\overline  W\,
  \label{tracedgravitino}
\end{align}
This gives with (\ref{hatdeltah}), and using the on-shell value (\ref{feh})
\begin{equation}
  \hat{\delta }h^0 + \ft12a^{-1} P_R\gamma ^\mu \Sigma _\mu \approx \frac{3}{\sqrt{2}a}\nabla _{\bar \alpha }\overline{W}\bar \epsilon \chi ^{\bar \alpha }=\nabla _{\bar \alpha }h_G^0 \delta z^{\bar\alpha} \,.
 \label{hatdelhW}
\end{equation}
The last term is the expected result; it is the covariant transformation of the on-shell value of $h^0$.

\section{Synopsis}
\label{sec:conclusions}

The natural geometric setting for chiral multiplets in ${\cal N}=1,~D=4$ supersymmetry or supergravity is that of a K\"ahler manifold.  In \cite{Freedman:2016qnq} we developed a formulation of global supersymmetric theories  that is manifestly covariant under holomorphic diffeomorphisms  of the target space.  In this paper we extend this covariant approach to supergravity, and present a covariant treatment, which includes the auxiliary fields  of  chiral multiplets.\footnote{There are earlier treatments of supergravity with auxiliary fields \cite{Kugo:1982mr,Kugo:1982cu} that are not fully covariant.}

We follow the superconformal approach to supergravity, which includes a compensator multiplet and therefore begins with a set of $n+1$ chiral multiplets $X^I, \Omega^I, \hat F^I$.  The notation $\hat F^I$ indicates that the usual auxiliary fields $F^I$ are modified so that they transform as  a vector under diffeomorphisms. The $X^I$ are holomorphic coordinates of an $n+1$ dimensional conformal K\"ahler manifold whose metric $N_{I\bar J}(X,\bar X)$ is homogeneous.  The superconformal group includes a chiral symmetry called the $T$ symmetry,  and we define supersymmetry transformations that are covariant under  diffeomorphisms and covariant derivatives that include the composite $T$-connection.  The covariant formulation leads to simplified actions and transformation rules.

The physical theory, which is invariant under Poincar\'e supersymmetry, contains $n$ chiral multiplets $z^\alpha,\chi^\alpha, h^\alpha$, which are defined in terms of the superconformal components in (\ref{Confvariables}).  The passage from
 superconformal to Poincar\'e requires gauge-fixing conditions for symmetries of the superconformal algebra that are not part of the  Poincar\'e  subalgebra.  The most important of these are the dilatation, $T$-symmetry, and the $S$-supersymmetry.
 The covariant supersymmetry transformations can be expressed in terms of the physical fields, but this is not quite enough.  The gauge-fixed action is invariant only if the gauge-fixing conditions are maintained, and this requires certain compensating transformations.  In the end we define covariant supersymmetry transformations of the physical fields that require only the compensating transformation for $S$-supersymmetry.

While the process described above is somewhat involved, the resulting passage from superconformal to the super-Poincar\'e theory is very simple.  The  final off-shell \Poincare theory contains $n$ the expected $h^\alpha$ auxiliary fields, which are auxiliary fields of the physical chiral multiplets, plus the field $h^0$  whose role is the same as the $S+\rmi P$ auxiliary field of the old minimal formalism. The SUSY transformations for scalars, fermions and auxiliary fields, all explicitly covariant, can be found in \eqref{delz}, \eqref{hatdelchi} and \eqref{hatdeltah}. We also derived the auxiliary field action in detail that is relevant for auxiliary fields, and discussed the relation between its on- and -off-shell forms.
One possible application of our framework would be the investigation of possible supergravity theories with novel non-linear realization of supersymmetry. The canonical example follows from a nilpotency condition $\Phi^2 = 0$ on a single superfield, which expresses the scalar field as a fermion bilinear. Other constraints involving e.g.~multiple chiral fields or the supergravity multiplet itself can be addressed in a similar fashion, with different relations between the components of these multiplets. We hope that our covariant approach provides a fruitful starting point for such investigations.

\medskip
\section*{Acknowledgements}

\noindent The authors warmly thank Renata Kallosh for suggesting this topic and for participating in the early stages of our research. We acknowledge hospitality of the Department of Physics of Stanford University during the visit in which this work was initiated, and of the GGI institute in Firenze during the workshop `Supergravity: what next?'.

The research of DZF is partly supported by the US National Science Foundation grant NSF PHY-1620045.
The work of AVP is supported in part by the Interuniversity Attraction Poles Programme
initiated by the Belgian Science Policy (P7/37).
This work is supported in part by the COST Action MP1210 `The
String Theory Universe'.

\bibliography{supergravity}
\bibliographystyle{toine}

\end{document}